\documentclass{article}
\usepackage{arxiv}

\usepackage[utf8]{inputenc} % allow utf-8 input
\usepackage[T1]{fontenc}    % use 8-bit T1 fonts
\usepackage{hyperref}       % hyperlinks
\usepackage{url}            % simple URL typesetting
\usepackage{booktabs}       % professional-quality tables
\usepackage{amsfonts}       % blackboard math symbols
\usepackage{nicefrac}       % compact symbols for 1/2, etc.
\usepackage{microtype}      % microtypography
\usepackage{lipsum}		% Can be removed after putting your text content
\usepackage{graphicx}
\usepackage{natbib}
\usepackage{doi}

\usepackage{todonotes}
\usepackage{changes}

\usepackage{wrapfig}
\usepackage{amsmath,amssymb,bm}
\usepackage{subcaption,booktabs}
\newcommand{\E}{\mathbb{E}}

\usepackage{multirow}
\usepackage{makecell}
\usepackage{pifont}
\newcommand{\cmark}{\ding{51}}%
\newcommand{\xmark}{\ding{55}}%

\newtheorem{mydef}{Definition}
\presetkeys{todonotes}{inline}{}
\usepackage{enumitem}
\setlist[itemize]{leftmargin=*}

\title{A review of cooperation in multi-agent learning}

\author{ 
       Yali Du\\
        King's College London\\
        \texttt{yali.du@kcl.ac.uk} \\
 \And
        Joel Z. Leibo \\
        Google DeepMind\\
        \texttt{jzl@deepmind.com} \\
\And
        Usman Islam \\
        King's College London\\
        \texttt{usman.islam@kcl.ac.uk} \\
\AND
     Richard Willis \\
        King's College London\\
        \texttt{richard.willis@kcl.ac.uk} \\
 \And
        Peter Sunehag \\
        Google DeepMind\\
        	\texttt{sunehag@deepmind.com} \\
}
\date{}
\renewcommand{\shorttitle}{Cooperation in MAL: A Review}

\hypersetup{
pdftitle={A template for the arxiv style},
pdfsubject={q-bio.NC, q-bio.QM},
pdfauthor={David S.~Hippocampus, Elias D.~Striatum},
pdfkeywords={First keyword, Second keyword, More},
}

\begin{document}
\maketitle
\begin{abstract} 

Cooperation in multi-agent learning (MAL) is a topic at the intersection of numerous disciplines, including game theory, economics, social sciences, and evolutionary biology. Research in this area aims to understand both how agents can coordinate effectively when goals are aligned and how they may cooperate in settings where gains from working together are possible but possibilities for conflict abound. 
In this paper we provide an overview of the fundamental concepts, problem settings and algorithms of multi-agent learning. This encompasses reinforcement learning, multi-agent sequential decision-making, challenges associated with multi-agent cooperation, and a comprehensive review of recent progress, along with an evaluation of relevant metrics. Finally we discuss open challenges in the field with the aim of inspiring new avenues for research.
%The topic is gaining increasing attention, as the need for cooperative AI becomes more widely recognised and imperative. This is particularly relevant as more capable agents, utilising large foundation models, are being developed.
% This article delves into the algorithms and strategies that enable multiple agents to learn to collaborate in shared environments. 

\end{abstract}

% \yali{centralisation, or decentralisation for mixed motivation setting? norms, if inequity aversion is emergent?}

% % \added{
% In this paper, we will give a review of the concepts and fundamentals of multi-agent learning, including the basics of reinforcement learning, multi-agent sequential decision making,  the challenges for multi-agent cooperation and a review of progress along with evaluating metrics. We will also discuss open challenges and hopeful open up new research.

% e will also discuss the progress on evaluation metricooperative learning for self-interested agents; and 4) directions for future work.
% keywords can be removed

\section{Introduction}

{
% \rc
 % This might connect with the various other fields like game theory, economics, social science, evolutionary biology that it intersects a bit with. We also do not have to be complete for what is multi-agent RL like reviewing exploration methods or what other topics are not squarely focused around cooperation. Giving the reader an understanding of what has been done so far in cooperation provides the reader with the background for new research related to the growing interest in Cooperative AI.
%In the rapidly evolving domain of artificial intelligence (AI), cooperative 
%Multi-agent learning stands at the intersection of numerous disciplines. 

Cooperative multi-agent learning (MAL) delves into algorithms and strategies that allow multiple agents to learn how to collaborate, adapt, and make decisions in shared environments. As multi-agent systems become increasingly prevalent in our technologically driven world, the importance of ensuring effective and seamless cooperation between agents grows too.

Cooperative MAL naturally intersects with various other fields including economics \citep{zheng2021ai, johanson2022emergent} and evolutionary biology \citep{jaderberg2019human, duenez2023social}. Other concepts from the social sciences also play a large role such as communication, norms, and trust \citep{hertz2023beyond}.
Game theory provides a robust foundation for understanding  strategic interactions between agents including collaborative and non-collaborative decision-making \citep{shapley1953stochastic, littman1994markov}. Its mathematical formalism aligns with economic principles, and is especially useful when agents need to maximise shared utilities or when mechanisms are required to encourage cooperation in settings rife with potential conflicts.

While the broader field of MAL encompasses a wide range of topics, we aim to focus on its collaborative dimension. As the momentum around Cooperative AI grows (e.g.~\citep{dafoe20__open_problems_in_cooperative_ai}), it becomes imperative to offer readers a synthesised understanding of the area. The field has two main branches: team-based MAL (covered in Section \ref{sec:team-based}) and mixed-motive MAL ( covered in Section  \ref{sec:mixed-motive}). 

In team-based MAL, it is difficult to learn effectively coordinated joint policies because a single scalar reward signal provides the only feedback available on the activities of all agents on the team. Consider what happens when one agent takes a rewarding action while another agent act unhelpfully. The shared scalar reward cannot distinguish which agent's action was the one responsible for the reward. This makes credit-assignment difficult in this setting \citep{claus1998dynamics, foerster2018COMA, sunehag2018value}. 

In the mixed-motive setting there are individual rewards, which are easier to learn from. However, such games contain many sub-optimal equilibria, a fact which gives rise to social dilemmas----i.e.~situations where there is tension between individual and collective rationality \citep{rapoport1974prisoner}. In MAL, the game-theoretic notion of a social dilemma has been generalised to the spatially/temporally-extended complex behaviour learning setting \citep{leibo2017multi}. This area has seen the development of a vast array of techniques for enabling cooperation that more resembles what is seen in the human world and, therefore, intersects more with social science and evolutionary biology where the emergence of cooperation is an important topic of study \citep{duenez2023social}. 
For convenience, we use the term 'co-player' to describe other agents in team-based and mixed-motive settings, as opposed to 'opponent' in zero-sum settings.
% By traversing this interdisciplinary landscape, we hope to provide a comprehensive snapshot of the state-of-the-art, the methodologies in use, and the horizons yet to be explored. Whether you are a seasoned researcher or a newcomer, this review aims to equip you with a foundational grasp of the past, an appreciation of the present, and an anticipation for the future of cooperation in multi-agent learning.

% \paragraph{Structure of this paper} 
The structure of this paper is outlined as follows. Section \ref{sec:background} presents self-contained fundamental knowledge of multi-agent learning, including single agent and multi-agent RL, game theoretic formulations.
% problem formulations, basic solutions, and existing challenges. 
Section \ref{sec:team-based} considers cooperative systems with pure motivation. Section \ref{sec:mixed-motive} discuss the case where agents have mixed motivation. 
% We also discuss emergent cooperative behaviour. 
Section \ref{sec:eval} reviews the benchmarks and evaluation metrics. Section \ref{sec:conclusion} concludes with a discussion on challenges and open questions in the field.

\section{Background}\label{sec:background}
In this section, we provide the necessary background on  reinforcement learning, in both single- and multi-agent settings.
Specifically, we provide a brief overview of stochastic or Markov games, which are the most commonly used frameworks for describing the multi-agent learning setting.

\subsection{Single-agent Reinforcement learning}
Reinforcement Learning (RL) is a standard problem setting in ML consisting of an agent sequentially performing actions in an environment, leading to a change of observed state and a reward at each timestep. We first describe the single-agent setting using the Markov Decision Process (MDP) formalism and then extend this to the partially observable and multi-agent settings.

% \paragraph{Markov Decision Processes (MDP)}
\begin{mydef}
A Markov Decision Process is defined as a five-tuple $(S, A, P, r, \gamma)$ where: $S$ is a set of states, $A$ is a set of actions, $P(s' | s, a)$, the transition function, gives the probability that the environment transitions from state $s$ to state $s'$ given that the agent plays action $a$, and $r(s, a)$, the reward function, gives the expected reward obtained by the agent for performing action $a$ in state $s$. $\gamma \in [0, 1]$ is the discount factor.
\end{mydef}
Note that the transition and reward functions only depend on the current state and action, with no need for access to the history of the process. This is known as the Markov property. 
The Markov Decision Process (MDP) is a commonly used framework for modeling the decision-making process of an agent that has complete knowledge of the system state, denoted as $s$. In this framework, at each time step $t$, the agent selects an action $a_t$ based on the current state $s_t$. This action leads to a transition to a new state, $s_{t+1}$, which follows a probability distribution denoted as $\mathcal{P}\left(\cdot \mid s_t, a_t\right)$. Additionally, the agent receives an immediate reward, denoted as $R\left(s_t, a_t, s_{t+1}\right)$. The primary objective in solving an MDP is to find a policy $\pi: \mathcal{S} \rightarrow \Delta(\mathcal{A})$, which is a mapping from the state space $\mathcal{S}$ to a distribution over the action space $\mathcal{A}$. This policy, denoted as $a_t \sim \pi\left(\cdot \mid s_t\right)$, aims to maximise the expected sum of discounted rewards:
\begin{equation}
\mathbb{E}[\sum_{t \geq 0} \gamma^t R\left(s_t, a_t, s_{t+1}\right) \mid a_t \sim \pi(\cdot \mid s_t), s_0].    
\end{equation}
In this context, two essential functions are defined under policy $\pi$: the state-action function (Q-function) and the value function, expressed as:

% \begin{subequations}
% \begin{equation}
%     Q_\pi(s, a)  =\mathbb{E}[\sum_{t \geq 0} \gamma^t r(s_t, a_t, s_{t+1}) \mid a_t \sim \pi(\cdot \mid s_t), a_0=a, s_0=s], 
% \end{equation}
% \begin{equation}
%       V_\pi(s)  =\mathbb{E}[\sum_{t>0} \gamma^t r(s_t, a_t, s_{t+1}) \mid a_t \sim \pi(\cdot \mid s_t), s_0=s]
% \end{equation}
% \end{subequations}
% $$
\begin{align}
Q_\pi(s, a) & =\mathbb{E}[\sum_{t \geq 0} \gamma^t r(s_t, a_t, s_{t+1}) \mid a_t \sim \pi(\cdot \mid s_t), a_0=a, s_0=s], \\
V_\pi(s) & =\mathbb{E}[\sum_{t>0} \gamma^t r(s_t, a_t, s_{t+1}) \mid a_t \sim \pi(\cdot \mid s_t), s_0=s].
\end{align}
% $$
These functions quantify the expected cumulative rewards when starting from a specific state-action pair $\left(s, a\right)$ and state $s$, respectively. When referring to the functions associated with the optimal policy $\pi^*$, they are commonly known as the optimal Q-function and the optimal value function.

In many situations, it is too strong to assume the agent can see all the state all the time. Therefore an adjusted formalism can be employed known as a partially observable MDP (POMDP), defined as a 7-tuple $(S, A, P, R, \Omega, O, \gamma)$ where: $S$ is a set of states,  $A$ is a set of actions, and $P(s' | s, a)$ is the transition function giving the probability that the environment transitions from state $s$ to state $s'$ given that the agent plays action $a$. $r(s, a)$ is the reward function that gives the expected reward obtained by the agent for performing action $a$ in state $s$. $\Omega$ is a set of observations. $O(o | s', a)$, the observation function, gives the probability of observing $o \in \Omega$ given the reached state $s'$ and the action $a$. $\gamma \in [0, 1]$ is the discount factor. 
Different from the fully observable setting, at time $t$, the agent only observes $o$ from the environment with probability $O(o | s', a)$. 
% The objective remains the same.
Partial observations result in reduced sample efficiency since the agent needs more experience to perceive the full variety of possible state. Moreover, observations may be modelled in a noisy way (e.g. taking sensor noise into account), reducing stability and requiring even more experience to gain a reliable idea of the effect of actions on the environment (See \citet{cassandra1998exact} for more details on POMDP Model).
As we will see, these difficulties carry forward into the multi-agent partially observable case. 

RL algorithms generally follow two paradigms: value-iteration and policy-iteration. The former seeks to optimise a parameterised value or action-value function using an objective derived by dynamic programming and setting the policy to 
maximise the learned function, while the latter uses directly explore the policy space.
% Gradient Theorem to directly optimise the policy.

\paragraph{Value-based Methods}

% {\rc 
% Value-based RL methods are devised to find a good estimate of the state-action value function, namely, the optimal Q-function $Q_{\pi^*}$. The (approximate) optimal policy can then be extracted by taking the greedy action of the Q-function estimate. One of the most popular value-based algorithms is Q-learning \citep{watkins1992q}, where the agent maintains an estimate of the Q-value function $\hat{Q}(s, a)$. When transitioning from stateaction pair $(s, a)$ to next state $s^{\prime}$, the agent receives a payoff $r$ and updates the Q-function according to:
% $$
% \hat{Q}(s, a) \leftarrow(1-\alpha) \hat{Q}(s, a)+\alpha\left[r+\gamma \max _{a^{\prime}} \hat{Q}\left(s^{\prime}, a^{\prime}\right)\right]
% $$
% where $\alpha>0$ is the stepsize/learning rate. Under certain conditions on $\alpha, \mathrm{Q}$-learning can be proved to converge to the optimal Q-value function almost surely \citep{watkins1992q,szepesvari1999unified}, with discrete and finite state and action spaces. Moreover, when combined with neural networks for function approximation, deep Qlearning has achieved great empirical breakthroughs in human-level control applications (Mnih et al., 2015). Another popular on-policy value-based method is SARSA, whose convergence was established in Singh et al. (2000) for finite-space settings.
% }
Value-based reinforcement learning techniques aim to calculate an accurate approximation of the state-action value function, specifically the optimal Q-function denoted as $Q_{\pi^*}$. The approximate optimal policy can then be determined by selecting the action with the highest Q-function estimate. One well-known value-based algorithm is Q-learning \citep{watkins1992q}. In Q-learning, the agent maintains an estimate of the Q-value function denoted as $\hat{Q}(s, a)$. When the agent transitions from a state-action pair $(s, a)$ to the next state $s^{\prime}$, it receives a reward denoted as $r$ and updates the Q-function as follows:
\begin{equation}
    \hat{Q}(s, a) \leftarrow (1-\alpha) \hat{Q}(s, a) + \alpha [ r + \gamma \max_{a^{\prime}} \hat{Q}(s^{\prime}, a^{\prime}) ].
\end{equation}
% $$
% \hat{Q}(s, a) \leftarrow (1-\alpha) \hat{Q}(s, a) + \alpha [ r + \gamma \max_{a^{\prime}} \hat{Q}(s^{\prime}, a^{\prime}) ]
% $$
Here, $\alpha > 0$ represents the learning rate or step size. Under certain conditions on $\alpha$, Q-learning can be mathematically proven to converge to the optimal Q-value function almost surely \citep{watkins1992q,szepesvari1999unified}, with discrete and finite state and action spaces.
% with a high degree of certainty. This convergence has been demonstrated in prior research by Watkins in 1992 and Szepesvári in 1999, particularly when dealing with discrete and finite state and action spaces. 
Furthermore, when combined with neural networks for function approximation, deep Q-learning has shown remarkable empirical success in achieving human-level control in various applications, as demonstrated by  \citep{mnih2015human,mnih2013playing}.
Another notable on-policy value-based method is SARSA. Its convergence properties were established in a study by \citep{singh2000convergence}, particularly in settings with for finite-space settings.

% \todo{refer to  \citeauthor{zhang2021multi}}
% kaiqing

\paragraph{Policy-Based Methods}

Another category of reinforcement learning (RL) algorithms involves a direct exploration of the policy space, which typically utilise parameterised function approximators, namely approximating the policy by $\pi_\theta(\cdot \mid s)$ where $\theta$ denotes unknown parameters. Consequently, the straightforward approach of updating the parameter based on the gradient of long-term rewards has been implemented through the policy gradient (PG) method. The fundamental premise behind this concept is expressed as \citep{sutton1999policy}:

\begin{equation}
\nabla J(\theta) = \mathbb{E}_{a \sim \pi_\theta(\cdot \mid s), s \sim \eta_{\pi_\theta}(\cdot)}\left[Q_{\pi_\theta}(s, a) \nabla \log \pi_\theta(a \mid s)\right].
\end{equation}
Here, $J(\theta)$ represents the expected return, $Q_{\pi_\theta}$ is the Q-function under policy $\pi_\theta$, $\nabla \log \pi_\theta(a \mid s)$ denotes the policy's score function, and $\eta_{\pi_\theta}$ signifies the state occupancy measure, which can be either discounted or ergodic, under policy $\pi_\theta$. Various policy gradient methods, including REINFORCE \citep{williams1992simple}, and actor-critic algorithms \citep{konda1999actor}, have been introduced by estimating the gradient in various ways. This idea also applies to deterministic policies in continuous-action settings, and \citet{silver2014deterministic} derived the policy gradient for such cases. In addition to gradient-based methods, several other policy optimisation techniques have demonstrated excellent performance in numerous applications. These include PPO \citep{schulman2017proximal}, TRPO \citep{schulman2015trust}, and soft actor-critic \citep{haarnoja2018soft}.

% Compared to value-based methods, policy-based approaches offer more robust convergence guarantees, especially when using neural networks for function approximation. This makes them well-suited to handle extensive or even continuous state-action spaces, as demonstrated by previous research (Konda and Tsitsiklis, 2000; Yang et al., 2018; Zhang et al., 2019; Agarwal et al., 2019; Liu et al., 2019; Wang et al., 2019)."

\subsection{Multi-agent learning}
% \begin{wrapfigure}{r}{0.5\textwidth}
\begin{figure}
% {r}{0.5\textwidth}
% \vspace{-11pt}
\centering
\includegraphics[width=8.5cm]{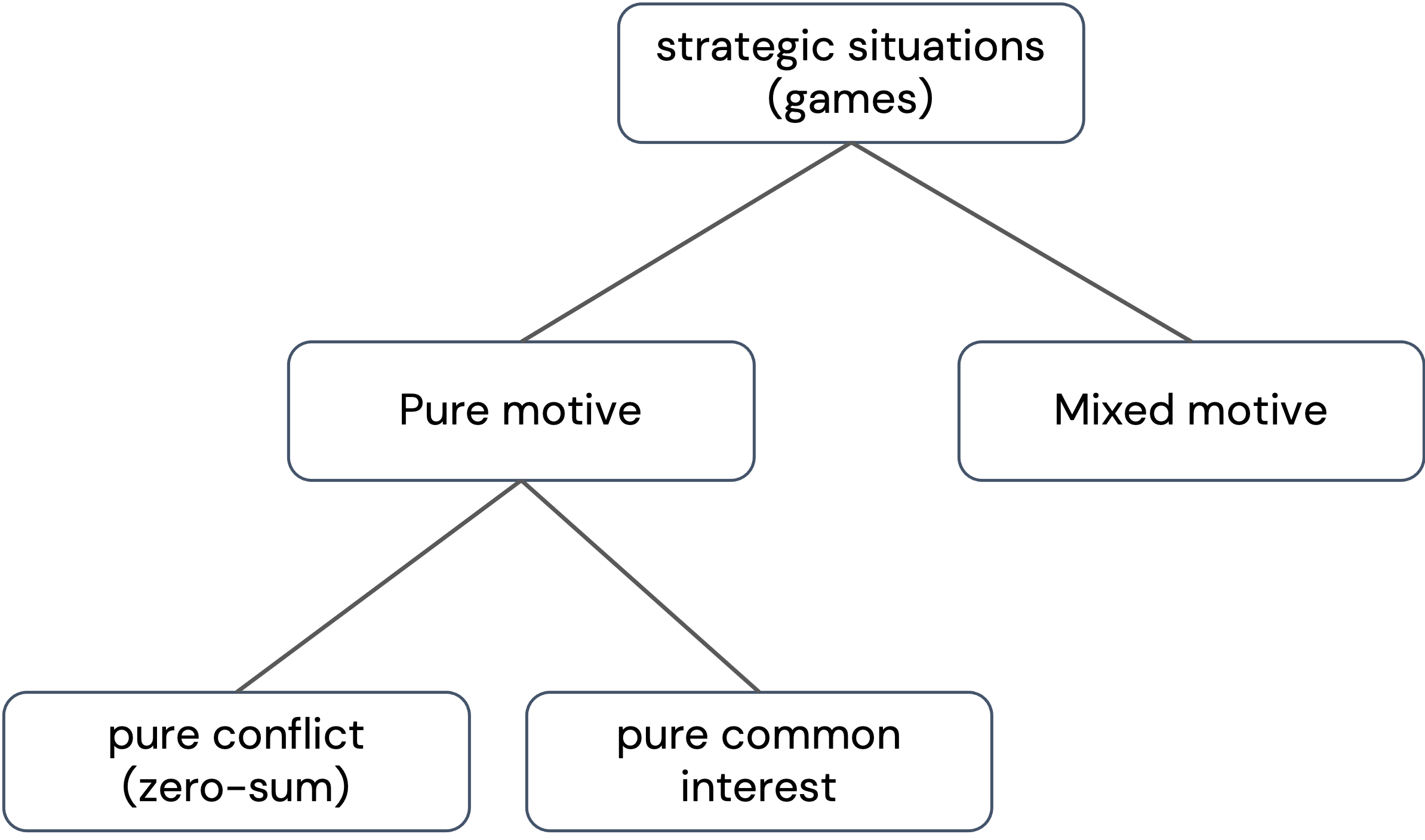}
    % \vspace{-20pt}
\caption{A taxonomy of multi-agent systems \citep{schelling1960strategy}. 
}
% \yali{revise to add more algorithms into it.} 

\label{fig:taxonomy}
% \end{wrapfigure}
\end{figure}

Multi-agent learning considers the more general situation with multiple interacting agents. 
The single-agent RL algorithm may not be applicable as this induces non-stationarity, i.e. the environment itself is changing from the perspective of each agent as the other agents' policies evolve. Accordingly, the problem is much harder and specific algorithms must be employed. 
% We give an overview of the cooperative case in this work but first we present the standard formalisms.
Figure \ref{fig:taxonomy} provides an taxnomy of multi-agent systems based on its utility structure. Multi-agent systems fall into the following categories:
\text{Cooperative}, where the agents share a common reward, 
 \text{Competitive}, where rewards are individual and sum to zero (i.e. one agent's gain is another's loss), 
\text{mixed-motive}, where rewards are individual and both cooperative and competitive motivations coexist.

% \paragraph{Markov Games}
\begin{mydef}
A Markov Game is defined by a set of $N$ agents (or players) and the following:
$S$ is a set of states, $A_i$ is a set of actions for agent $i \in \mathbb{N}=\{1,2,...,N\} $,
$P(s' | s, a)$, the transition function, gives the probability that the environment transitions from state $s$ to state $s'$ given that the agents play the joint action $\bm{a} := (a_1, ..., a_k)$,
$R_i(s, \bm{a})$, the reward function for agent $i$, gives the expected reward obtained by $i$ after the agents perform the joint action $a$ in state $s$,
$\gamma \in [0, 1]$ is the discount factor.
\end{mydef} 
% A Markov game (or stochastic game) is a multi-agent extension of MDP. Agent $i$ seeks to maximise the return $\sum_{t'=t}^{T}\gamma^{t' - t}r_{t'}^i$, where $r_{t'}^i$ is the reward obtained by $i$ at time $t'$.
{ Along the same spirit with POMDP, in Markov games, it is not always true to assume that the agent can see all the state all the time. Therefore an adjusted formalism can be employed known as a Partially Observable Markov Games (POMG). 
Players are unable to directly observe each state; rather, they obtain a partial observation of the state, which is defined by the observation function $o_i\in \mathbb{R}^d$, which is determined by an observation function $\mathcal{O}: \mathcal{S} \times \mathbb{N} \rightarrow \mathbb{R}^d$.
% applicable to S ×.
% Players
% cannot perceive each state directly, but instead receive their own  partial observation
% of the state
% , which is determined by the observation function O : S ×
% .
The partially-observed case has been predominant in the research of mixed-motive settings and is sometimes omitted to reduce the clutter in terminology. 
}
The framework of Markov games is a general umbrella of various multi-agent learning settings, namely cooperative, competitive and mixed-motive settings. 
\paragraph{Cooperative Setting}
In completely cooperative scenarios, all agents typically work under a unified reward function, symbolised as $R_1 = R_2 = ... = R_N = R$. This model is recognised as multi-agent MDPs \citep{boutilier1996planning,lauer2000algorithm} or team Markov games \citep{wang2002reinforcement}.

Given this framework, both the value function and Q-function are consistent across all agents. This uniformity allows the use of single-agent RL techniques, like the Q-learning update, when all agents function as a single decision entity. The apex of cooperation in this context aligns with a Nash equilibrium in the game's landscape.

In a variant setting, agents can possess distinct reward functions, possibly confidential to each agent \citep{zhang2018fully}. 
% Another emerging model for cooperative multi-agent reinforcement learning, which is slightly more encompassing than the universal reward model, is the team-average reward approach \citep{zhang2018fully}. Here, agents can possess distinct reward functions, possibly confidential to each agent. 
The cooperative objective is to enhance the enduring reward based on the average reward $\bar{R}= \frac{1}{N}\sum R^i(s,\bm{a},s')$
for any given $(s,\bm{a},s')$ set in $S \times A \times S$. 
This model, emphasising agent variation, counts the previously discussed model as a subset. It not only maintains agent privacy but also propels the creation of independent multi-agent RL algorithms \citep{zhang2018fully,qu2020scalable}. This heterogenity further demands integrating communication structures into multi-agent reinforcement learning and examining communication-efficient strategies.

\paragraph{Competitive Setting}

For competitive environments are typically represented as zero-sum Markov games. In essence, the collective rewards of all participants equal zero for every combination of $(s,a,s')$. The bulk of studies has predominantly centred on dual-agent competing with each other \citep{littman1994markov} since, in this scenario, one agent's gain directly corresponds to the other's loss. Apart from direct applications in gameplay \citep{littman1994markov,silver2016mastering,OpenAI_dota}, such zero-sum scenarios are also employed to advance robust learning, as the uncertainty that prohibits learning could be modelled as a fictitious opponent that has opposite interest with the robust learning agent.
% that hinders an agent's learning trajectory can be conceptualised as an ever-opposing fictional adversary within the game (citing )\yali{to refien}. 
The Nash equilibrium gives rise to a policy that's tailored for optimising rewards in the harshest scenarios.

% \todo{add ref}
% Fully competitive setting in MARL is typically modelled as zero-sum Markov games, namely, Pi∈N Ri(s,a,s′) = 0 for any (s,a,s′). For ease of algorithm design and analysis, most literature focused on two agents that compete against each other (Littman, 1994), where clearly the reward of one agent is exactly the loss of the other. In addition to direct
% 8
% applications to game-playing (Littman, 1994; Silver et al., 2017; OpenAI, 2017), zero-sum games also serve as a model for robust learning, since the uncertainty that impedes the learning process of the agent can be accounted for as a fictitious opponent in the game that is always against the agent (Jacobson, 1973; Bas ̧ar and Bernhard, 1995; Zhang et al., 2019). Therefore, the Nash equilibrium yields a robust policy that optimises the worst-case long-term reward.

\paragraph{Mixed Setting}
The mixed setting, often referred to as the general-sum game framework, doesn't place limitations on the objectives or interactions of agents \citep{hu2003nash,littman2001friend,littman1994markov}. These agents act based on their individual interests, and their rewards can sometimes be at odds with those of other agents. Foundational game theory concepts like Nash equilibrium \citep{bacsar1998dynamic} greatly shape the algorithms tailored for such a setting. Multi-agent problems can also be considered under scenarios combining both fully cooperative and competitive agents within this setting. An example would be two teams in a zero-sum competition, where agents within each team collaborate \citep{OpenAI_dota, jaderberg2019human}.
In this paper, we focus on the cooperation in both cooperative settings and mixed settings. 

\section{Challenges in Multi-agent Cooperation}
% \section{Challenges in Multi-agent \replaced{Cooperation }{Learning} }

% \paragraph{MARL Challenges}
% The problem of cooperation in multi-agent systems (MAS) has become increasingly important in AI, game theory and real applications. 
% The use of reinforcement learning (RL), in particular, has attracted recent attention \citep{vinyals2019grandmaster,silver2016mastering}. Using RL as a means
% of achieving coordinated behaviour is attractive because of its
% generality and robustness.
In contrast to single-agent reinforcement learning, where the agent's objective is to efficiently maximise long-term returns, the learning objectives in multi-agent reinforcement learning are sometimes less clearly-defined. The field debates whether MAL is to be understood as a question or as an answer \citep{shoham2007if}. In fact both motivations exist simultaneously.
There are two main lines of research: one focuses on maximising joint values, described as Team Markov Games or Multi-agent MDPs, and the other seeks to find common ground among agents and promote their social welfare or avoid social dilemmas.
% , modelled by general-sum or mixed motivation games. 
These scenarios are often modelled through mixed-motive (general sum) games.
In Section \ref{sec:challenge-team}, we summarise below the key challenges for multi-agent cooperation in Team Markov Games. 
While mixed motivation games share a similar set of challenges with Team Markov Games, they also involve new challenges arising from the self-motivated nature of the agents, which are discussed in Section \ref{sec:challenge-mix}. 

% \textcolor{red}{JZL: Why do we use the names ``Team Markov Games'' and ``mixed-motive games'' for the two subfields here? What’s the point of highlighting the fact that the setting is in a Markov game for the team case but not highlighting the same thing for the mixed-motive case? We could refer to ``mixed-motive Markov games'' for consistency with the name we use for the team case. Or alternatively, why are we highlighting that they are Markov games in the subfield names at all? We could just refer to the `team-based setting' and the `mixed-motive setting'. There's no need for either subfield name to include the word Markov.}

\subsection{Shared challenges appearing in both team and mixed-motive settings}

\paragraph{Non-stationarity and scalability in the number of agents} 
%\paragraph{Scalability}
Multi-agent systems are highly non-stationary, since any agent's policy improvement is experienced by other agents in the system  as a change in the distribution of their experience \citep{foerster2017stabilising, lowe2017multi, leibo2019autocurricula}. From each agent's perspective, the environment involves all the other agents and their own interactions. 

When you increase the number of agents you also increase the size of the joint action space exponentially. This may lead to exploding training times for algorithms that attempt to model the joint action space directly. A naive way to sidestep this problem is to use independent and decentralized learning algorithms for each agent \citep{claus1998dynamics}. However, performance degrades in this case because non-stationarity is not taken into account \citep{foerster2017stabilising}. Exploration, a key part of any RL scheme, also ostensibly becomes more difficult since the choice of actions to explore now needs to be coordinated in joint action space, rather than each agent naively adding independent noise to its action as one might do in single-agent reinforcement learning \citep{bard2020hanabi}. On the other hand, \cite{leibo19__malthusian_reinforcement_learning} argued that the non-stationarities induced by increasing population sizes can actually be a blessing in disguise since under certain conditions a rising number of players may induce a kind of  consistently-directed exploration called ``exploration by exploitation'' which would be hard to motivate otherwise due to the presence of training plateaus and attractive local optima (see also \citet{sunehag2023diversity}).

%Furthermore, it is well-established in the literature that addressing the general Dec-POMDP problem is computationally challenging, as noted by \citep{bernstein2002complexity,oliehoek2016concise}.
%This partial information exacerbates the challenges associated with non-stationarity, as it is hard to accurately recover the  behaviour of the other players’ underlying policies, thereby amplifying the non-stationarity perceived by individual agents.

\paragraph{Cooperating with Novel Partners (Generalisable Social behaviour)}

This challenge concerns developing agents that are capable of cooperating with novel individuals who they never encountered during prior training \citep{rahman2021towards, hu2020other, leibo2021scalable, jaderberg2019human, carroll2019utility}. It is also called the problem of ``ad hoc teamwork'' \citep{stone2010ad}, but that name is somewhat misleading since it implies a team sports metaphor which is too limited in practice to cover much of the work in this area which includes a wide range of efforts to study and improve generalization to novel social situations. Some work in this area is concerned with pure common-interest situations where agents are aligned with strangers on the goal but still must effectively coordinate to achieve it. Other work in this area is concerned with mixed-motive situations where agents meet strangers with whom they may not even be aligned on the overall goal \citep{leibo2021scalable, agapiou2022melting}. To be robust to novel agents, an agent needs to be able to engage in reciprocal cooperation with like-minded partners, while remaining robust to exploitation attempts from others.
In order to do this, the agents need a broad skill set, including the social skills to determine who they can trust to stick to their agreements or reciprocate, as well as the ability to flexibly adapt when they encounter conventions different to those previously experienced \citep{dafoe20__open_problems_in_cooperative_ai}.
They also need flexibility to adapt when they encounter different conventions to those previously experienced.
A core challenge when developing these capabilities in an agent is to expose them to sufficiently diverse behaviours during training \citep{strouse2021collaborating, madhushani2023heterogeneous} (see also \citet{balduzzi2019open} for discussion of the analogous need for training diversity in  situations of pure conflict).

In this setting, coordination can be difficult to achieve because it is not possible to rely on prior knowledge to anticipate how a novel player will act. One example approach is to develop policies that are effective against a predetermined set of teammate \text{types}. An agent then classifies each novel teammate based upon their action history and selects a suitable behavioural policy for their type \citep{strouse2021collaborating,lou2023pecan,li2023cooperative}.
% \textcolor{red}{(JZL: Did you have a specific reference in mind for this? Please add it.)}
A zero-sum version of this idea appears in the OPRE algorithm of \cite{vezhnevets2020options}, which was extended to the mixed-motive setting in \cite{leibo2021scalable} and \cite{agapiou2022melting}.

\subsection{Team Markov Games}\label{sec:challenge-team}

\paragraph{Credit Assignment}

 % \todo{any better word for this}
In single-agent RL, credit assignment is the problem of distinguishing which past actions contributed significantly to the current reward, or equivalently how much the current action will contribute to future rewards, which is sometimes termed temporal credit assignment or reward redistribution \citep{sutton2018reinforcement}.
% \citep{fang2019curriculum,zhang2023interpretable,dai2022diversity}. 
% \textcolor{red}{JZL: all three references here are single-agent RL papers. That's probably too many to include here. They distract the reader from MAL. In fact there's not really any need to emphasize the connection to credit assignment in single-agent RL so much here in this section (a single reference to a textbook would be enough). The credit-assignment issue in MAL is different from the usual one in single-agent RL. We need to make that clear to the reader.}

Credit assignment is difficult in multi-agent team scenarios where agents strive to optimise a joint reward. The problem stems from the inability of decentralized agents to precisely determine their own contribution to the reward. It may lead to  complacency in some agents over time (an issue sometimes called the lazy agent problem \citep{sunehag2018value}) and ineffective coordination. Moreover, reinforcement learning agents generally have a hard time differentiating between their teammates' exploration behaviour and random environmental factors \citep{claus1998dynamics}. Partial observability also exacerbates credit assignment issues since it makes it difficult for an agent to get reliable information about co-player behaviour and how it may impact their shared reward. 
% {\rc While a rich line of work studied how to distribute rewards among team players \cite{jeon2022maser,wang2020shapley} in dense reward setting, 
% sparse and delayed rewards in the multi-agent reinforcement learning (MARL) context remain less explored. Recent attempts leverage transformer to perform global reward decomposition, into local rewards \citep{chen2023stas,she2022agent}, but success is only witnessed in small scale scenarios and lacks interpretability. Especially it is worth pointing out that the episodic reward setting where rewards are only seen at the end of the episodic is difficult and remain less understood.

While a rich body of research has focused on distributing rewards among team players in dense reward settings \citep{jeon2022maser, wang2020shapley}, the area of sparse and delayed rewards in the context of multi-agent reinforcement learning (MARL) remains relatively unexplored. Recent efforts have utilized transformers for global reward decomposition into local rewards \citep{chen2023stas, she2022agent}. However, success has been limited to small-scale scenarios and often lacks interpretability. It is particularly noteworthy that understanding the episodic reward setting, where rewards are only observed at the end of an episode, is challenging and not well-understood.

\subsection{Mixed-motive games}\label{sec:challenge-mix}
% \todo{include discussion on challenges on mixed-motive case. refer to Leibo2017aaams}
% \yali{@Joel, please double check. }

\paragraph{Heterogeneous incentives}
%Mixed-motive games exhibit the same challenges as team games, as detailed above.
In mixed-motive situations there is the additional complication that the agents desire different outcomes from each other.
This means that an agent cannot rely on their partners to act in their best interests.
For example, a game of football is a team game, each player wants their team to win, and the challenge is to determine and enact the most successful strategy taking into account the capabilities of their teammates and their opponents.
If the most successful strategy were known, each teammate would be content to play their corresponding role.
Now, suppose there is an additional financial incentive given to the players for each goal that they score.
This makes the game a mixed-motive game, as each player now has a preference to win the game, as before, but also to personally score goals \citep{koster2020model}.
Whereas without the financial incentive, a teammate can be trusted to pass the ball if this is more likely to result in a goal, now they cannot be fully relied upon to do so, because they might prefer to be the player who takes the shot.

Here, at least the personal incentives still broadly align with the collective objective, as scoring goals will increase the chance that the team will win.
In some mixed-motive games, called social dilemmas, individual incentives actually conflict with the group's objective \citep{rapoport1974prisoner}.
For example, suppose the unscrupulous manager of the opposing team were to offer a far larger financial incentive to the first player to score an own-goal.
Now, what is best for the players (scoring an own-goal before anyone else) conflicts with what is best for their team (winning).
Even if the strategy that maximises the probability of the team winning were known, the players would have incentives to deviate from it.
In these situations, if each player attempts to do what is best for themselves, this can lead to poor outcomes for the group.
Notable examples are \text{free-riders} who benefit from \text{public goods} without contributing their share \citep{hughes2018inequity}, and the \text{tragedy of the commons} where a \text{common pool resource} is degraded by unrestrained consumption \citep{perolat2017multi}.

\paragraph{Collective good}
% Due to the difference in preferences, it can be hard for the agents to agree on which outcome to target.
Consider the case where a user is training a group of agents and is deciding what to use as a performance metric.
% Even here, it can be difficult to determine what outcomes are most desirable.
Suppose that the user wishes to maximise the collective good or social welfare.
Due to differences in individual preferences, it can be difficult to understand what is best for the collective.
First, there are many possible notions of the collective good, such as the utilitarian, which measures the sum of all individual agent rewards, or  \cite{rawls71__a_theory_of_justice}'s metric, which measures the reward of the worst-off agent.
Second, even with a clear understanding of the chosen concept of social welfare, there is also the issue of fairness to take into account.
Even though a given outcome may maximise the sum of player rewards, if it results in an unfair distribution of rewards, this may not be desirable.
Therefore another metric, such as the Gini coefficient \citep{ceriani12__the_origins_of_the_gini_index} that measure inequality could be used in tandem \citep{perolat2017multi}.
Once desirable performance has been specified, it is still non-trivial to achieve it.
Due to the misalignment between individual incentives, simply training each agent to optimise their own rewards is unlikely to result in good outcomes \citep{leibo2017multi,mckee2020social}.
Additional mechanisms are typically required to achieve cooperation in these settings.

\section{Cooperation in Team Games} \label{sec:team-based}
%\section{Cooperation with Pure Motivation} \label{sec:team-based}
% ith  agents sharing common payoffs is the most studied domain. 

In the fully cooperative situation, all agents typically work under a unified reward function, symbolised as $R_1 = R_2 = ... = R_N = R$. Due to its relevance to single-agent tasks, it is arguably the most studied domain.
In this section, we first provide a high-level explanation of the taxonomy, which is constructed based on the learning paradigm and type of policies, and then of the representative algorithms for team game settings. 
Table \ref{tab:algs-team} summarises recent popular algorithms for solving common-payoff games.

% {\rc 
% Compared to the single-agent case, the information structure of MARL, namely, who knows what at the training and execution, is more involved. 
% For example, in the framework of Markov games, it suffices to observe the instantaneous state st, in order for each agent to make decisions, since the local policy pi maps from s to A. 

% Learning schemes resulting from various information structures lead to many variations of algorithms.

% The extreme case is the aforementioned independent learning scheme, which assumes the observability of only the local action and reward, and suffers from non-convergence in general \citep{tan1993multi}.
% }
% \newpage 
\begin{table}[t!]
\caption{Summary of representative algortihms for MARL with common payoffs. VB and PG represent learning types, with VB standing for Value-based and PG for Policy Gradient.
% \yali{@Usman, pls replace Alg name with citations. merge columsn of VB and PB, remove evolution based, } Value-based/ \\ Policy Gradient
}
    \label{tab:algs-team}
    \centering
\vspace{3mm}
\begin{tabular}{ c|c|c|c|cc} 
\hline
% Algorithm & \multicolumn{1}{|p{2cm}|}{\centering centralised/ \\ Decentralised \\ Critic} & \multicolumn{1}{|p{2cm}|}{\centering centralised/ \\ Decentralised \\ Actors} & \multicolumn{1}{|p{1.5cm}|}{\centering Value-based/ \\ Policy Gradient} & \multicolumn{1}{|p{1.5cm}|}{\centering Policy- \\ based}  & Communication \\
Algorithm & \multicolumn{1}{|p{2cm}|}{\centering Centralised/ \\ Decentralised \\ Critic} & \multicolumn{1}{|p{2cm}|}{\centering Centralised/ \\ Decentralised \\ Actors} & \multicolumn{1}{|p{2cm}|}{\centering Learning}  & Communication \\
\hline
MADDPG \citep{lowe2017MADDPG} & C & D & PG  & \xmark \\ 
COMA \citep{foerster2018COMA} & C & D & PG  &  \xmark \\  
LIIR\citep{du2019learning} & C & D & PG  & \xmark \\
GridNet \citep{han2019grid} & C & C & PG  & \cmark \\

MA2C \citep{iqbal2019actor} & C & D & PG &  \xmark \\
HAPPO/HATRPO \citep{kuba2021trust} & C & D & PG & \xmark \\
MAPPO \citep{yu2022surprising} & C & D & PG & \xmark \\  
VDN \citep{sunehag2018value} & C & D & VB  & \xmark \\
QMIX \citep{rashid2018QMIX} & C & D & VB  &  \xmark \\
QTRAN \citep{son2019qtran} & C & D & VB  &  \xmark \\
Qatten \citep{yang2020qatten} & C & D & VB  & \xmark \\
QPLEX \citep{wang2020qplex} & C & D & VB & \xmark \\
SHAQ \citep{wang2020shapley} & C & D & VB & \xmark \\
FMA-FQI  \citep{wang2021towards}& C & D & VB & \xmark \\

\hline
RIAL/DIAL \citep{foerster2016dial} & C & D & VB & \cmark \\
CommNet \citep{sukhbaatar2016commNet} & C & D & PG & \cmark \\
ATOC \citep{jiang2018learning}  & C & D & PG  & \cmark \\
TarMAC \citep{das2019tarmac} & C & D & PG  &  \cmark \\
IC3Net \citep{singh2019ic3net} & C & D & PG  &  \cmark \\
SchedNet \citep{kim2019SchedNet} & C & D & PG  & \cmark \\ 
DCG \citep{bohmer2020dcg}  & C & D & VB & \cmark \\
DGN \citep{jiang2020graph}  & C & D & VB &  \cmark \\
\hline
Networked AC \citep{zhang2018fully} & D & D & PG &  \cmark \\
% MF-Q/MF-AC \citep{yang2018mean} & D & D & VB/PG &  \cmark \\
NeurComm \citep{chu2020multiagent} & D & D & PG &  \cmark \\
% \citep{gupta2020networked} & C & D & \xmark & \cmark & \cmark \\
% NetES & C & D & \xmark & \xmark &  \cmark \\
% Networked Actor-Critic  & C & D & \xmark & \cmark &  \cmark \\
\hline
\end{tabular}
\end{table}
\subsection{Learning Paradigm}
\label{sec:paradigm}
In comparison to the single-agent scenario, multi-agent reinforcement learning (MARL) introduces a more complex information structure, which determines who has knowledge during training and execution. For instance, in the context of Markov games, it is sufficient for each agent to observe the current state $s$ to make decisions, as the local policy $\pi$ maps from states to actions.

Various learning approaches stemming from different information structures result in numerous algorithmic variations. An extreme case is the independent learning scheme, where agents only have visibility of local actions and rewards. This scheme generally encounters convergence issues \citep{tan1993multi}.
Independent learning is a straightforward method that adapts single-agent RL algorithms for multi-agent environments, where each agent operates as a separate learner. In such a framework, the actions of other agents are seen as environmental factors. This concept was initially conceptualised in \cite{tan1993multi}, where the Q-learning algorithm was adapted for this context, leading to what is known as Independent Q-Learning (IQL). The primary hurdle with IQL lies in its non-stationarity, given that actions from individual agents aiming for local objectives influence environmental transitions.

In order to address the challenge of handling partial information as described earlier, a substantial body of research has operated under the assumption of a central controller. This central entity is responsible for gathering data such as collective actions, shared rewards, and joint observations, and is even design policies for all participating agents.
However, it is important to note that in most practical applications, a centralised controller is not readily available, except in cases where access to a simulator is easily attainable, such as in video games and robotics \citep{peng2017BicNet,han2019grid}

To strike a balance between decentralised control and non-stationarity, the popular learning scheme of \text{centralised-learning-decentralised-execution }  (CTDE) has emerged. This concept originated from research in planning for the partially observed setting, specifically, Dec-POMDP \citep{oliehoek2016concise}.
% An alternative framework is centralised Training with Decentralised Execution (CTDE),
In CTDE framework, the critic (i.e. value or action-value function neural network) takes all agents' observations and actions into account but each agent's policy only takes its own observation into account. Here the actors are decentralised, leading to fast exploration, but training of the critic incorporates all agents in order to capture the non-stationarity and coordination of behaviours. 
When the same team rewards are distributed among all agents, a single critic model suffices. However, when rewards are localised and private, each agent should have its critic model for training.

% Some methods train one critic shared across all agents while others train one for each agent, depending on if the agents have the shared or individual rewards. 
% Here we consider the former.

\subsection{Solution approaches} \label{sec:sub:solution-for-team-game}

% Depending on the information strucutre of the actors, whether the 

\paragraph{Policy-based Methods}
% \yali{algorithm name to be removed, to make it consistent in flow }
% \paragraph{centralised Actor}

The first class of algorithms employs the policy-iteration paradigm discussed earlier. The difference is in the structure of the policy/policies across agents.

According to the Policy Gradient Theorem, the objective for policy updates in the single-agent case is
\begin{equation}
    g = \E_\pi[Q(s, a)\nabla_\theta \ln\pi_\theta(a | \tau)],
\end{equation} 
where $Q(s, a)$ is the critic. 
While $\pi_\theta(a | \tau)$ can be modelled as the joint policies of individual $\pi_i(a | \tau)$, this approach necessitates a centralised controller, and the research in this pathway usually aim to achieve high scalability to number of actors \citep{han2019grid,peng2017BicNet}.
For instance, GridNet \citep{han2019grid} represents the state information as a grid feature map and employs convolutional neural networks as the policy network to achieve scalable control of an arbitrary number of agents.
On the other hand, BicNet \citep{peng2017BicNet} models the interdependencies of agents through the use of bi-directional RNN. Although it operates in a centralised manner, it offers flexibility in controlling varying numbers of agents. This method is said to employ a \text{centralised actor} since there is only one policy and this takes all agents' states into account in order to produce all actions.

For decentralised actors, it can be extended to:
\begin{equation}
    g_i = \E_\pi[Q(s, \bm{a})\nabla_\theta \ln\pi_i(a_i | \tau_i)],
\end{equation}
% \[g_i = \E_\pi[Q(s, \bm{a})\nabla_\theta \ln\pi_i(a_i | \tau_i)]\]
for agent $i$. To reduce variance, we can subtract from the critic a \text{baseline} function which is independent of the action of that agent (this will not change the optimal parameters). This is similar to the single-agent case but here the baseline can incorporate the other agents' actions since the policy is trained in a decentralised way:
\begin{equation}\label{}
g_i = \E_\pi[(Q(s, \bm{a}) - b(s, \bm{a}_{-i}))\nabla_\theta \ln\pi_i(a_i | \tau_i)].    
\end{equation}
% \textbf{COMA} 
Note that we have estimated the action-value function (A.K.A. Q-function) using the reward obtained using the current action and the value function of the next state. 
Representative algorithms include MAPPO \citep{yu2022surprising}, HAPPO \citep{kuba2021trust}, COMA \citep{foerster2018COMA} and LIIR \citep{du2019learning}.

% \textbf{MADDPG}
In cases where agents' actions are continuous, deterministic policies are employed for agents. According to single-agent DDPG \citep{lillicrap2015continuous}, the deterministic policy gradient update is
\begin{equation}
  \nabla_\theta J(\theta) = \E_{s}[\nabla_\theta\mu_\theta(a | s)\nabla_\theta Q^\mu(s, a)].
\end{equation}
Here a deterministic policy $\mu_\theta$ is used instead of $\pi_\theta$. The multi-agent extension for agent $i$ is given as follows 
\begin{equation}
 \nabla_{\theta_i} J(\mu_i) = \E_{\bm{x}, a}[\nabla_{\theta_i}\mu_i(a_i | s_i)\nabla_{a_i} Q_i^\mu(\bm{x}, a_1, ..., a_N)],
\end{equation}
where the critic is trained by standard TD learning. This is the basis for Multi-agent DDPG \citep{lowe2017MADDPG}. Note that each agent has its own critic, meaning that this can model cooperative, competitive or mixed behaviours. We assume the agents know each others actions, but if this is not the case the authors present a method to infer these, by maximising the log probability of other agents' actions with an entropy regulariser. MADDPG also solves the Markov game problem, but the limitation for such algorithm is difficulties in proving convergence.
% and a limitation is that it is difficult to prove convergence.

% \textbf{MAPPO}
MAPPO \citep{yu2022surprising} is a straightforward extension of PPO to the CTDE multi-agent case, with the caveat that agents must share policy parameters. It solves the Dec-POMDP problem so we assume shared rewards and value functions which take in all the agents' actions as opposed to the policies which are decentralised.
% \textbf{HAPPO and HATRPO}
Whereas MAPPO requires agents to share parameters, HAPPO and HATRPO \citep{kuba2021trust} overcome this limitation. They are multi-agent extensions of PPO and TRPO respectively, justified theoretically by a multi-agent advantage decomposition result. Namely, the joint advantage function can be decomposed into the sum of agents' local advantages. Since parameter sharing is not necessary, these methods solve the Markov game problem (where competitive and mixed behaviour is possible).

The standard baseline function is the value function $V(s^t)$, leading to the \text{advantage} function $Q(s^t, \bm{a}) - V(s^t) = r + \gamma V(s^{t+1}) - V(s^t)$ being substituted into the policy gradient objective. However, this choice only takes global rewards into account so it does not do a good job of assigning credit for rewards to specific agents. Counterfactual Multi-agent Policy Gradients (COMA) \citep{foerster2018COMA} instead uses a counterfactual baseline, i.e. the Q-function with the given player marginalised out:
\begin{equation}
    b(s, \bm{a}_{-i})) := \sum_{a'_i}{\pi_i(a'_i | s_i)Q(s, (a'_i, \bm{a}_{-i}))}.
\end{equation}
The adjusted advantage function now compares the Q-value using all agents' actions to one using only the other agents' actions. The marginalisation is done over the current estimate of the agent's policy. Each agent's policy objective now encodes more specific information about the agent's contribution. 
% COMA is built on the Markov game formalism.

% \textbf{LIIR}  
In the similar vein, Learning Individual Intrinsic Reward (LIIR) \citep{du2019learning} adds a learned intrinsic reward function $r_i^{in}(s_i, a_i)$ per agent to the standard extrinsic reward $r^{ex}$ from the environment. $r_i^{in}(s_i, a_i)$ is parameterised by $\eta_i$ and takes in the given agent's state and action. It is trained to maximise the extrinsic (i.e. standard) discounted return $J_ex$ so that it is in line with the overall MARL problem, while using the experience of the specific agent. 
The proxy reward is defined as $r^{proxy} := r^{ex} + r^{in}$ and the proxy discounted return $J^{proxy}$ is then used to train the policy parameters $\theta_i$. Note that since $r^{in}$ is trained to maximise the extrinsic return, the proxy return preserves the MARL objective for policy updates. However, each agent's return now reflects its own specific experience as well, leading to increased diversity between agents. This is a CTDE method for fully cooperative multi-agent systems and the overall update is done as a bilevel optimisation using meta-gradient.
% One challenge in team-based MARL is learning diversified behaviour for the agents even though there is one common reward.

\paragraph{Value-based Methods}
In the following we review some popular methods that focus on the critic and learn using CTDE. A centralised critic is one which is shared among the agents and takes in all state and actions. This lacks scalability as it must be trained on the joint action space. On the other hand, a decentralised critic for a given agent only takes in the observations and actions of that agent. Using decentralised critics does not account for the non-stationarity inherent in multi-agent systems.
Therefore, a better alternative to these two approaches is \text{value factorisation}, where we start with decentralised critics and pass the outputs into a \text{mixing network} whose output represents a combined (centralised) critic network. This allows fast training based on only localised experience while also taking agent interactions into account. We would like that the optimiser of the centralised critic $Q_{tot}$ also individually optimises each decentralised critic $Q_i$. This is known as the Individual-global Maximisation (IGM) constraint:
\begin{equation}
  \text{argmax}_{\bm{a}}Q_{tot}(\bm{\tau}, \bm{a}) = (\text{argmax}_{a_1}Q_1(\tau_1, a_1), ..., \text{argmax}_{a_n}Q_n(\tau_n, a_n)).  
\end{equation}
The methods differ in how exactly they factorise the value function. We will consider \textbf{linear} and \textbf{nonlinear factorisation approaches}.
% \paragraph{Linear factorisation}
% \textbf{VDN} 
Value Decomposition Network (VDN) \citep{sunehag2018value} simply sums the decentralised critics to produce the full critic so that:
\begin{equation}
    Q_{tot}(\bm{\tau}, \bm{a}) = \sum_i{Q_i(\tau_i, a_i)}.
\end{equation}
TD learning is then performed on the full critic so that the agents are not trained using their own specific rewards. This method is scalable due to the simple summation and it satisfies the IGM constraint. However, linear factorisations are a limited representation and there is no global convergence guarantee for value-based learning.

% \textbf{QMIX}
Although the IGM constraint is satisfied by VDN, the summation factorisation is a rather strong  requirement. QMIX \citep{rashid2018QMIX} notes that it is sufficient to have a global argmax on $Q_{tot}$ that yields the same result as a set of individual argmax operations on the decentralised critics. 
% QMIX, Rashid et al. in 2018, notes that it is sufficient for a global argmax on the total Q-value to produce the same result as a set of individual argmax operations on decentralized critics.
Therefore it makes a weaker requirement, namely that $Q_{tot}$ is monotonically increasing with each individual $Q_i$. In other words, the partial derivative of $Q_{tot}$ with respect to $Q_i$ is non-negative:
\begin{equation}
    \frac{\partial Q_{tot}}{\partial Q_i} \geq 0, \forall{i}.
\end{equation}
The mixing network is not a summation but a full neural network with the constraint that the weights must be non-negative, thus ensuring the above criterion. This allows for a much more complex representation of the centralised critic while allowing decentralised policies to be extracted using linear-time individual argmax operations (i.e. over individual actions rather than the joint action).

% \textbf{QTRAN}
QTRAN \citep{son2019qtran} builds on the above two methods but presents an even weaker assumption on the mixing network that is nevertheless sufficient to satisfy the IGM constraint. The assumption requires the non-negativity of an expression involving $Q_{tot}$, $\sum_i{Q_i}$ as well as the (state-)value function.
The authors note that this property is also \text{necessary} under an affine transformation. The end result is a more general mixing network that can model a wider class of MARL problems than VDN and QMIX.
% \textbf{Qatten}
Qatten \citep{yang2020qatten} employs a multi-head attention based mixing network for $Q_{tot}$, allowing the critic to explicitly measure the importance of each individual $Q_i$ to $Q_{tot}$.
% \textbf{FMA-FQI} \citep{wang2021towards} 
FMA-FQI \citep{wang2021towards} formalise a multi-agent fitted Q-iteration framework to analyse factorised multi-agent Q-learning. Within this framework, they explore linear value factorisation and uncover that multi-agent Q-learning with this straightforward decomposition implicitly achieves a robust counterfactual credit assignment, though it might not converge in certain scenarios. 
% Further analysis indicates that on-policy training or a richer joint value function can enhance its local or global convergence properties. To validate their theoretical insights, they provide an empirical evaluation of cutting-edge deep multi-agent Q-learning algorithms on instructive examples and an extensive range of StarCraft II unit micromanagement tasks.

% \yali{to check}
% TODO: is this a MARL algorithm or a theoretical framework to analyse other MARL algorithms?
% \peter{Peter: I believe this work builds upon Fan et. al. 2019 that is an attempt at analysing DQN by performing a simplification that turns the experience replay into neural fitted Q-iteration and that paper claims that they are able to prove some convergence results in this case. The paper mentioned here tries to take the same approach to Value-Decomposition-Networks and assume that the value-function is decomposable and then show that convergence can (under some further assumptions) be achieved by doing Q-learning on the VDN form of value function. I am not sure to what extent these technically quite complex papers are believed to be correct.
% }

% \paragraph{Nonlinear factorisation}
% \textbf{QPLEX}
QPLEX \citep{wang2020qplex} uses a dueling mixing network based on dueling DQN \citep{wang2016dueling}. This involves expressing the action-value functions (both joint and for individual agents) in terms of advantage functions and state-value functions. Specifically, we have:
% \begin{subequations}
    \begin{align}
    % \end{aligned}
        Q_{tot}(\bm{\tau}, \bm{a}) &= V_{tot}(\bm{\tau}) + A_{tot}(\bm{\tau}, \bm{a}),  \\  
    % \end{equation}
    % \begin{equation}    
    Q_i(\tau_i, a_i) & = V_i(\tau_i) + A_i(\tau_i, a_i)),    \end{align}
% \end{subequations}
where $V_{tot}(\bm{\tau}) = \max_{\bm{a'}}{Q_{tot}(\bm{\tau}, \bm{a'})}$ (and similarly for the individual value functions). The idea is to move the IGM constraint from the Q-functions to the advantage functions:
\begin{equation}
  \text{argmax}_{\bm{a}}A_{tot}(\bm{\tau}, \bm{a}) = (\text{argmax}_{a_1}A_1(\tau_1, a_1), ..., \text{argmax}_{a_n}A_n(\tau_n, a_n)).
\end{equation}

The benefit of this is that the constraint can be directly realised by limiting the value of advantage functions. It is known as the advantage-based IGM constraint and this is an equivalent transformation since the state-value terms do not affect the action selection. To obtain the final factorisation, we need to substitute the individual state-values and advantages into the expression for $Q_{tot}$. Since the state-value function does not take in actions, we can simply set $V_{tot}(\bm{\tau}) = \sum_i{V_i(\tau_i)}$. For the advantages, we will need importance weights since it involves the actions:
\begin{equation}
  A_{tot}(\bm{\tau}, \bm{a}) = \sum_i{\lambda_i(\bm{\tau}, \bm{a})A_i(\tau_i, a_i)},
\end{equation}
where the importance weights satisfy $\lambda_i(\bm{\tau}, \bm{a}) > 0$. Overall, we then have the full factorisation as:
\begin{equation}
  Q_{tot}(\bm{\tau}, \bm{a}) = \sum_i{V_i(\tau_i)} + \sum_i{\lambda_i(\bm{\tau}, \bm{a})A_i(\tau_i, a_i)}.
\end{equation}
This is trained end-to-end with the importance weights being learned using multi-head attention.

Several limitations of CTDE are worth noting. First, $Q(s, \bm{a})$ represents only an estimation of the global return and is trained using value iteration. This potentially limits its scalability to large scale problems. Second, CTDE does not adapt to the specific characteristics or requirements of individual agents. In the following section, we discuss how to distinguish between individual agents' contributions in the global return.

\subsection{Communication}
% \todo{to refine, add formulations of how agent communicate.}

% \yali{add discuion on LLM communication here. new work.}
% \subsection{Communication-based Methods}
% One notable limitation of the fully observable critic model is the potential communication overhead. Specifically, as the agent count increases, gathering all state/action data for a critic can become unfeasible due to constraints in communication bandwidth and memory.
A usual premise in decentralised policies is that agents are allowed to communicate with neighbors to a given extent. Such communication has been instrumental in enhancing exploration, maximising reward, and diversifying solutions in intricate optimisation simulations \citep{barkoczi2016social}, as well as in human studies \citep{lazer2007network}.
Section \ref{sec:sub:solution-for-team-game} has reviewed many non-communicative techniques prioritising stabilised training through centralised value assessment.

% \citep{chu2020multiagent} shows that communicate with neighbors contribute to enlarge agent's observation in a delayed manner.

% Effective communication is vital in Multi-Agent Reinforcement Learning (MARL) to discern the interdependencies of agents' actions, especially for independent execution. Such communication has been instrumental in enhancing exploration, maximising reward, and diversifying solutions in intricate optimisation simulations \citep{barkoczi2016social}, as well as in human studies \citep{lazer2007network}.
% \added{\usman{Is this paragraph needed? Seems like it's just describing the previous sections.}
% MARL methodologies can be grouped based on their communication strategies. 
% First, there are non-communicative techniques prioritising stabilised training through centralised value assessment. Notable methods in this category include COMA \citep{foerster2018COMA}, MADDPG \citep{lowe2017MADDPG}, and LIIR \citep{du2019learning} which are discused in Section \ref{sec:sub:solution-for-team-game}.

One line of work employs fixed communication topology for both training and execution.
Works like RIAL/DIAL \citep{foerster2016dial} leverage shared policy fingerprints among agents, while CommNet \citep{sukhbaatar2016commNet} utilises distinct communication channels combined with average pooling to assimilate information. Approaches such as DCG \citep{bohmer2020dcg}, ATOC \citep{jiang2018learning} and DGN \citep{jiang2020graph} identify neighboring agents through the $K$-nearest neighbor mechanism. In contrast, studies like GridNet \citep{han2019grid} employ convolution techniques in their policy networks to indirectly harness neighboring information. Research like CommNet \citep{sukhbaatar2016commNet} delves into communication structures among various learning agents and primarily focuses on static topologies. Similarly, Networked MARL methods like MA2C \citep{chu2020multiagent}, Networked Actor-Critic \citep{zhang2018fully} and \cite{gupta2020networked} consider scenarios where communication is confined to connected neighborhoods in interconnected systems.

Recently, attention-based communication methods form the third group, focusing on discerning communication recipients. Examples include VAIN \citep{hoshen2017vain}, which adapts the approach from CommNet to introduce an attention vector for agent selection.  Techniques like TarMAC \citep{das2019tarmac}, IC3Net \citep{singh2019ic3net}, MAGIC \citep{niu2021multi} and SchedNet \citep{kim2019SchedNet} employ binary attention mechanisms to govern communication instances.
% ATOC \citep{jiang2018learning} assigns communication weights among agents. \usman{ATOC is in both categories}

Another special line of work studied decentralised learning with networked agents where agents are allowed to communicate with their neighbors over a prescribed network, such as traffic grid and mobile sensor networks.
\cite{zhang2018fully} studied actor critic algorithm with guarantee to convergence to a consensus. 
In consensus algorithms or in the fully observable critic model, it's presupposed that agents can share their observations, actions, or rewards with their counterparts. The aspiration here is that, equipped with information from their peers, they can collectively discern the optimal policy. \citet{zhang2019decentralized} provide a review of recent advances on decentralised training for networked agents.

\subsection{Credit assignment}
\label{sec:credit}
% In MARL, we have the additional question of which agents contributed significantly to the rewards. 
% This compounds the credit assignment problem which concerns how to distinguish agents contributions to the rewards. 
% and may lead to poor coordination as the agents do not have a sound idea of other agents' important actions.

% In multi-agent scenarios where agents strive to optimise a collective reward, or when it's feasible to simplify the multi-agent challenge to a single-agent one, standard RL algorithms might not achieve the global optimal outcome. This limitation stems from the agents' inability to precisely determine their reward contributions, leading to potential complacency in some agents over time.
Credit assignment problem  concerns how to distinguish agents contributions to the rewards, while only a shared team rewards available, and agents do not have a sound idea of other agent's contributions.
%  represents only an estimation of the global return and is trained using value iteration. This potentially limits its scalability to large scale problems
% Current cooperative MARL techniques mainly address determining an agent's role in the collective team achievement, predicated on the availability of frequent global team rewards.
The approaches to credit assignment can vary, whether they are indirect or direct methods. 
Indirect methods as seen in previous work \citep{wang2022shaq,wang2020shapley,foerster2018COMA}, where the collective state-action value is seen as a combination of each agent's state-action value, subsequently allocating the communal global rewards based on individual actions.
% \added{
For example, ShapleyQ \citep{wang2022shaq,wang2020shapley} present a cooperative game-theoretical framework, extending the Shapley value to Markov games, termed the Shapley Q-value. This Shapley Q-value provide individualised critic for based on each agent's individual contribution, offering a contrast to the shared critic approach.
The indirect strategies often grapple with constraints in their representational power, especially in continuous action domains, while
Direct methods grapple with discerning the individual impacts of separate agent actions on the collective rewards \citep{wolpert2001optimal}. 
% or direct methods like LIIR  \citep{du2019learning } that calculate intrinsic rewards \citep{wolpert2001optimal}.
 % using specific reward benchmarks
LIIR \citep{du2019learning} firstly introduced intrinsic motivation in multi-agent team games to encourage the diverse behaviour. To regulate the diverse behaviours to be reward-seeking, \cite{du2019learning} presented a bilevel programming approach to guide the training of intrinsic reward generator.
% In the similar vein, a line of work employs \citep{chen2023ljir,ma2022elign,zheng2021episodic,mguni2021ligs,jeon2022maser}
Sparse and delayed rewards in the multi-agent reinforcement learning (MARL) context remain less explored. Recent attempts leverage transformer to perform global reward decomposition, into local rewards \citep{chen2023stas,she2022agent}.

% \yali{add some paper on sparse reward marl, like cmae, wait for ziyan}

% {\rc  When the reward is shared among all agents, only one critic model is required; however, in the case of private local reward, each agent needs to train a local critic model for itself. }

 % \subsection{Other topics}
% \paragraph{Credit Assignment in Actor-Critic Methods}

% \subsection{Summary}
% Next section we will review the multi-agent cooperation problem with individualised objectives.

% \subsection{}
% \yali{todo}
% \yali{Capture flag: Some math details can be added}
% \begin{center}
% \end{center}
% \subsection{Benchmark environment}
% \begin{itemize}
%     \item MPE
%     \item SMAC
%     \item MA-Mujoco
% \end{itemize}

\subsection{Team-Based Coordination with Novel Partners}
% \subsection{Ad Hoc Team Work} 
% yali: I felt that Ad Hoc Team work is a less clear domain with the problem not clearly defined. So i decided to not mention them. 
The aim of team-based coordination with novel partners, also known as zero-shot coordination, is to construct AI agents that can coordinate in pure common-interest scenarios with novel partners they have not seen before, such as humans or other AI players.  
While autonomous agents are trained with a given set of partners on completing some tasks, one challenge is to reason about the best way to collaborate with other agents and people without prior coordination, with applications seen in service robots, team sports and autonomous driving. Recent advances were mainly seen in resolving games such as Hanabi \citep{hu2020other,bard2020hanabi} and Overcooked \citep{carroll2019utility,lou2023pecan}, as well as in models of non-verbal communication \citep{bullard2020exploring}.

Learning algorithms aiming to facilitate zero-shot coordination at test time typically either depend on prior knowledge of environmental symmetries \citep{hu2020other} or rely on training with a maximally diverse set of co-players \citep{carroll2019utility, lupu2021trajectory,zhao2021maximum}. For instance, in an example of the latter approach, \cite{strouse2021collaborating} operate as an online evolutionary algorithm, continuously adjusting policy parameters and executing policy substitutions within the population to which its learning agents train to best respond. Likewise \cite{lupu2021trajectory} and \cite{zhao2021maximum} introduce additional auxiliary objective to improve the diversity of the population of partner policies it trains with.

\section{Cooperation with Mixed Motivation}\label{sec:mixed-motive}

% \section{mixed-motive settings}\label{sec:mixed-motive}

% \begin{figure}[t!]
% % \begin{wrapfigure}{r}{0.5\textwidth}
%     \centering
%     \includegraphics[width=0.5\linewidth]{figs/venn_diagram.pdf}
%     \caption{Social dilemma inequalities are satisfied in blue regions.}
%     \label{fig:venn_diagram}
% % \end{wrapfigure}
% \end{figure}

\begin{table}[t!] \label{tab:social-dilemma}
  \centering
  \caption{\label{table:sd} Social Dilemmas}
\vspace{3mm}

 \begin{subtable}{0.2\linewidth}
  % \color{blue}
    \centering
    \begin{tabular}{c|cc}
      & $C$ & $D$\\
      \hline
      $C$ & $R,R$ & $S,T$\\
      $D$ & $T,S$ & $P,P$\\
    \end{tabular}
    \subcaption{\label{table:generic_ssd} Social Dilemmas}
  \end{subtable}
  \begin{subtable}{0.2\linewidth}
    \centering
    \begin{tabular}{c|cc}
      & $C$ & $D$\\
      \hline
      $C$ & $3,3$ & $0,4$\\
      $D$ & $4,0$ & $1,1$\\
    \end{tabular}
    \subcaption{\label{table:sd_a}Prisoner's Dilemma}
  \end{subtable}
  \begin{subtable}{0.2\linewidth}
    \centering
    \begin{tabular}{c|cc}
      & $C$ & $D$\\
      \hline
      $C$ & $3,3$ & $1,4$\\
      $D$ & $4,1$ & $0,0$\\
    \end{tabular}
    \subcaption{\label{table:sd_b}Chicken}
  \end{subtable}
  \begin{subtable}{0.2\linewidth}
  % \color{blue}
    \centering
    \begin{tabular}{c|cc}
      & $C$ & $D$\\
      \hline
      $C$ & $4,4$ & $0,3$\\
      $D$ & $3,0$ & $1,1$\\
    \end{tabular}
    \subcaption{\label{table:sd_c}Stag Hunt}
  \end{subtable}
  % \begin{subtable}{0.2\linewidth}
  % % \color{blue}
  %   \centering
  %   \begin{tabular}{c|cc}
  %     & $C$ & $D$\\
  %     \hline
  %     $C$ & $10,7$ & $0,0$\\
  %     $D$ & $0,0$ & $7,10$\\
  %   \end{tabular}
  %   \subcaption{\label{table:battle_of_sex}Battle of Sexes}
  % \end{subtable}
   
\end{table}

% \rich{
% \begin{itemize}
% \item credit assignment is now individual?
% \item Include this paper? \url{https://arxiv.org/pdf/2102.12307.pdf}
% \end{itemize}
% }

This section will focus on the topic of mixed motivation, as shown in Figure \ref{fig:taxonomy}.
Unlike pure common interest, where the focus is on solving practical problems such as sensor networks or a fleet of warehouse robots that jointly achieve higher common payoffs, the emphasis in mixed motivation also includes individual performance.

% In the field of mixed motivation, applications can be found in areas such as mechanism design, and modeling and understanding decision-making processes in social systems.
% This can include studying the behaviour of individuals or animals and how they interact in a system.
% The process of modeling involves defining theories and models that reflect our understanding of the real world, and then using these models to make predictions about how social systems behave. Models can be differential equation models, iterated prisoner's dilemma models, etc.
% The ultimate goal is to capture and understand the phenomenon under examination.

\subsection{Social dilemmas}
Central to the study of cooperation arising from mixed motivations is the concept of social dilemma: a situation where there is tension between individual and collective rationality \citep{rapoport1974prisoner}. In a social dilemma agents may be conflicted between playing strategies for the good of all players (cooperating) and playing selfish but often individually rational strategies (defecting). More specifically, consider a two-player matrix game with two actions per player, interpreted as a cooperate action, $C$ and a defect action, $D$, respectively. 
Table \ref{table:generic_ssd} shows three matrix games which are commonly considered to be canonical social dilemmas in the literature \citep{macy2002learning,leibo2017multi}. 
There are four relevant payoffs here:
% the reward for mutual cooperation $R$, punishment for mutual defection $P$, temptation reward for a player defecting while the other cooperates $T$, and sucker punishment for a player cooperating while the other defects $S$.

\begin{itemize}
    \item \emph{Reward} ($R$) of mutual cooperation.
    \item \emph{Punishment} ($P$) arising from mutual defection.
    \item \emph{Temptation} ($T$) the outcome for the player who defects while their co-player cooperates.
    \item $S$ \emph{Sucker} ($S$) the outcome for the player who cooperates while their co-player defects.
\end{itemize}

% The payoff matrix for this game can therefore be presented as in Table \ref{table:generic_ssd}. 
This game is a social dilemma when the payoffs satisfy the \text{matrix game social dilemma} inequalities given in Table~\ref{table:mgsd_inequality}:
% \begin{align}\label{test}
%     & R > P \text{ (mutual cooperation $>$ mutual defection)}\\
%     & R > S  \text{(mutual cooperation $>$ being exploited)} \\
%     & 2R > T + S \text{ (sum of payoffs is greater for mutual cooperation than for exploitation)} \\
%     & \text{Either} \textit{greed}: T > R or \textit{fear}: P > S
% \end{align}

% \begin{subequations}
% \label{eq:social_dilemma_ineq}
%     \begin{align}
%     & R > P \text{ (mutual cooperation $>$ mutual defection)}\\
%     & R > S  \text{ (mutual cooperation $>$ being exploited)} \\
%     & 2R > T + S \text{ (sum of payoffs is greater for mutual cooperation than for exploitation)} \\
%     & \text{Either} \textit{ greed}: T > R  \text{ or } \text{ fear}: P > S
% \end{align}
% \end{subequations}

\begin{table}[t!]
\caption{\label{table:mgsd_inequality}Matrix game social dilemma inequalities}
\centering
\vspace{3mm}
\begin{tabular}{llp{10cm}}
\toprule
 & Inequality & Comment\\
\midrule
1 & \(R>P\) & Players prefer mutual cooperation over mutual defection.\\
2 & \(R>S\) & Players prefer mutual cooperation over unilateral cooperation.\\
3 & \(2R>T+S\) & Players prefer mutual cooperation over an equal probability of unilateral cooperation and defection.\\
 & At least one of: & \\
4a & \(T>R\) & Players prefer unilateral defection to mutual cooperation, which is known as \text{greed}.\\
4b & \(P>S\) & Players prefer mutual defection to unilateral cooperation, which is known as \text{fear}.\\
\bottomrule
\end{tabular}
\end{table}

While matrix game social dilemmas have been widely applied in social science, economics and biology, they have several shortcomings as models of social dilemmas in real life \citep{leibo2017multi}.
First, real-life dilemmas are stateful and temporally extended, instead of being stateless and one-shot.
Furthermore, cooperativeness may not be binary here, and an agent could display behaviour on a spectrum of cooperativeness, which may vary over time.
This means that players will react to an ongoing pattern of play by other players in a more complicated environment.
In addition, the initiation and effects of cooperate or defect behaviours may not occur simultaneously, and their effects as when player 1 starts executing a temporally extended strategy, player 2 may observe this and react accordingly.
Players may also only have partial information of the state of the environment. Finally, and most importantly, in complex environments one must learn not just which strategic choice to take as an ``atomic'' unit but rather must instead learn a whole policy to implement whatever choice they make. The dynamics of learning to implement one's choices also affect the outcomes \citep{hertz2023beyond}.
Although some of these additions can be modelled with repeated play, continuous action spaces and extended-form games, a natural choice is to use Markov games.

\paragraph{Sequential Social Dilemmas}
The above issues motivate the idea of a \text{sequential social dilemma} (SSD).
We now consider Markov games instead of matrix games and policy spaces instead of action spaces. Specifically, we policy sets $\Pi^C$ and $\Pi^D$ that implement cooperative and defecting policies respectively.
The players can choose policies $\pi^C \in \Pi^C$ or $\pi^D \in \Pi^D$ from these.
As for payoffs, notice that this new situation calls for considering the long-term sum of rewards of a particular policy given the other player's policy.
We therefore use the \text{expected} long-term rewards, defined analogously to the payoffs in the previous section.
First, we define player $i$'s value function:
\begin{equation}
V_i^{\pi}(s_0) = \mathbb{E}_{a_t \sim \pi(O(s_t)), {s_{t+1} \sim T(s_t, a_t)}}{\sum_{t=0}^{\infty}{\gamma^{t}r_i(s_t, a_t)}},
\end{equation}
where $\pi = (\pi_1, \pi_2)$ and $a_t$ is the joint action at time $t$. 
Now we can define $R$, $P$, $T$ and $S$ as functions of the state $s \in S$:
\begin{subequations}
    \begin{equation}
        R(s) := V_1^{\pi^C, \pi^C}(s) = V_2^{\pi^C, \pi^C}(s),
    \end{equation}
    \begin{equation}
    P(s) := V_1^{\pi^D, \pi^D}(s) = V_2^{\pi^D, \pi^D}(s),      
    \end{equation}
    \begin{equation}
    T(s) := V_1^{\pi^D, \pi^C}(s) = V_2^{\pi^C, \pi^D}(s),      
    \end{equation}
    \begin{equation}
    S(s) := V_1^{\pi^C, \pi^D}(s) = V_2^{\pi^D, \pi^C}(s).      
    \end{equation}
\end{subequations}

The game is now a sequential social dilemma if the social dilemma inequalities (Table~\ref{table:mgsd_inequality}) hold for $R(s)$, $P(s)$, $T(s)$ and $S(s)$.
In practice, the expected long-term payoffs are estimated by fixing the policies of the agents and averaging over multiple simulations, a technique known as \text{empirical game-theoretic analysis} \citep{walsh02__analyzing_complex_strategic_interactions_in_multiagent_systems,wellman06__methods_for_empirical_gametheoretic_analysis,tuyls20__bounds_and_dynamics_for_empirical_game_theoretic_analysis,viqueira20__improved_algorithms_for_learning_equilibria_in_simulationbased_games}.
We now present a formal definition.

\begin{mydef}[Sequential Social Dilemma \citep{leibo2017multi}] 
A sequential social dilemma is a tuple $\left(\mathcal{M}, \Pi^C, \Pi^D\right)$.
$\mathcal{M}=$ a Markov game.
$\Pi^C=$ set of cooperative policies.
$\Pi^D=$ set of defecting policies.
Consider the empirical payoff matrix $(R(s), P(s), S(s), T(s))$,
induced by policies $\left(\pi^C \in \Pi^C, \pi^D \in \Pi^D\right)$ via the payoffs defined above.
$\left(\mathcal{M}, \Pi^C, \Pi^D\right)$ is an SSD if its empirical payoff matrix $(R(s), P(s), S(s), T(s))$ satisfies the social dilemma inequalities in Table \ref{table:mgsd_inequality}. 
\end{mydef}

The generalisation to more than two players is as follows.
An $N$-player \text{sequential social dilemma} is a tuple $(\mathcal{M}, \Pi = \Pi_c \sqcup \Pi_d)$ of a Markov game and two disjoint sets of policies, said to implement cooperation and defection respectively, satisfying the following properties. Consider the strategy profile $(\pi^1_c, \dots, \pi^\ell_c, \pi^1_d, \dots, \pi^m_d) \in \Pi_c^\ell \times \Pi_d^m$ with $\ell + m = N$. We shall denote the average payoff for the cooperating policies by $R_c(\ell)$ and for the defecting policies by $R_d(\ell)$.

We say that $(\mathcal{M}, \Pi)$ is a sequential social dilemma iff the following hold:
\begin{enumerate}
    \item Mutual cooperation is preferred over mutual defection: $R_c(N) > R_d(0)$. \iffalse In the language of Schelling, there exists some coalition size $k > 0$ that is better than Nash. The regional social dilemma definition has $R_c(i) > R_d(j)$ for $i > j$. \fi
    
    \item Mutual cooperation is preferred to being exploited by defectors: $R_c(N) > R_c(0)$. \iffalse In other words, cooperation has a positive internalities. The social dilemma definition has $R_c(i) > R_d(j)$ for some $i > j$. \fi
    
    \item Either the \text{fear} property, the \text{greed} property, or both:\begin{itemize}
        \item Fear: mutual defection is preferred to being exploited. $R_d(i) > R_c(i)$ for sufficiently small $i$. 
        \item Greed: exploiting a cooperator is preferred to mutual cooperation. $R_d(i) > R_c(i)$ for sufficiently large $i$.
        \end{itemize}
\end{enumerate}

A sequential social dilemma is \text{intertemporal} if the choice to defect is optimal in the short-term.
More precisely, consider an individual $i$ and an arbitrary set of policies for the rest of the group.
Given a starting state, for all $k$ sufficiently small, the policy $\pi^i_k \in \Pi$ with maximum return in the next $k$ steps is a defecting policy. There is thus a tension between short-term individual incentives and the long-term collective interest.

\paragraph{Schelling diagrams}
A Schelling diagram is a game representation which highlights interdependencies between agents, showing how the choices of others shape one's own incentives \citep{schelling1973hockey}.
It represents games with any number of players $N \ge 2$, and assumes that each individual faces a binary choice.
Thus it is said to be a model of binary choice with externalities. We refer to one choice as \text{cooperation} ($C$), and the other as \text{defection} ($D$).
In a game governed by a Schelling diagram, the payoffs from choosing either option are driven only by one's own choice and the number of other individuals that chose the $C$ option.
This means the effect of one's choice is influenced by the cumulative effect of externalities arising from the choices of others.

A \text{Schelling diagram} plots the curves $R_c(\ell + 1)$ and $R_d(\ell)$, as shown in Figure \ref{fig:models_of_dm}.
Intuitively, the diagram displays the two possible payoffs to the $N^{\textrm{th}}$ player given that $\ell$ of the remaining players elect to cooperate and the rest defect.
The minimum number of players who must cooperate in order for each player cooperating to do better than a defector does when all players defect is called the minimum viable coalition size. A social dilemma with a larger minimum viable coalition size generally requires more coordination to resolve than a social dilemma with a smaller minimum viable coalition size.
%In other words, it is the minimum coalition size that is worthwhile.
% This is known as a \emph{minimum viable coalition}.

\paragraph{Asymmetrical dilemmas}
In reality, agents typically have individual differences in their reward structures, affordances (actions), and sensors (observations), which lead to asymmetric interactions. For example, in ecology \citep{sunehag2019reinforcement} and economics \citep{zheng2021economist, johanson2022emergent}, agents are seen as belonging to different species or as playing different social roles. Such role-dependent differences in capabilities and tastes create opportunities for cooperation that differ fundamentally from symmetric scenarios. In particular, they make it possible to realize overall welfare gains from trade.
% For example in ecological simulations with three species in a food chain \citep{sunehag2019flocking}, one can consistently see the emergence of symbiotic relations ships between the prey at the bottom of the food chain and the apex predators, where the prey seek shelter from their predator, which the apex predator wants to attract and feed on.
% The predator in the middle develops flocking behaviours to corner prey while presenting an apex predator with a confusing array of moving targets in varying directions. 
The formal definitions above apply to symmetrical games, where the rewards to a player depend only on their own policy, and the combination of policies chosen by their co-players.
An attempt to generalise the definition of a social dilemma has been made by \citet{willis2023JAAMAS}.

\begin{figure}[t!]
    \centering
    \includegraphics[width=0.8\linewidth]{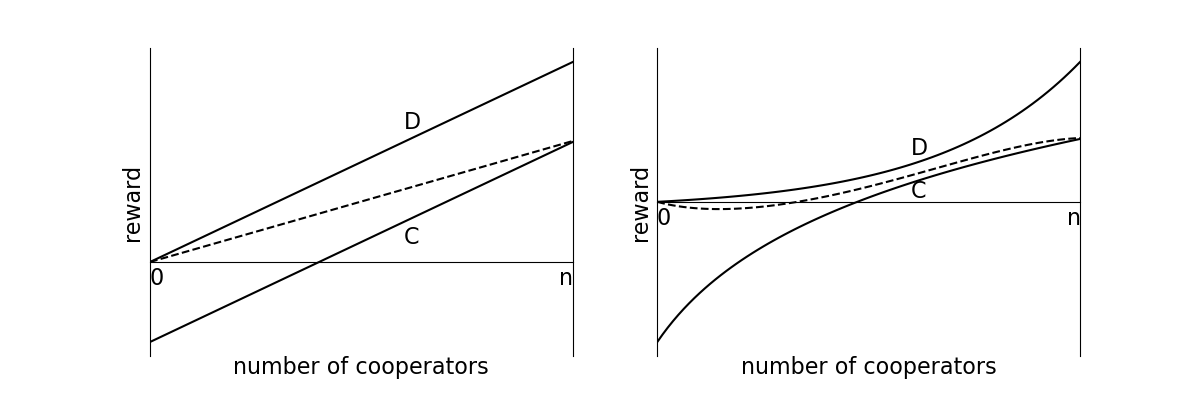}
    \caption{
    For a population of size $n$, the Schelling diagram shows the payoff for an $n+1$ agent choosing to either defect or cooperate. The dotted line shows the average reward of the population.
    } 
    \label{fig:models_of_dm}
\end{figure}

\subsection{Cooperation in sequential social dilemmas}
Numerous methods have been proposed to develop agents who exhibit good cooperation in sequential social dilemmas.
These methods have been introduced to achieve better group performance, often measured as the sum of reward for all agents.
In this section, we refer to the social welfare, $SW$, as the mean episode reward for every agent:
\begin{equation}
% \nonumber
SW(\bm{R}) = \frac{1}{n} \sum_i R_i.
\end{equation}
We discuss alternative notions of collective good in Section~\ref{sec:metrics}.

A common approach to induce cooperation is to use \text{reward shaping}, whereby an agent is provided with intrinsic reward in addition to its extrinsic game reward.
These additional rewards serve to motivate particular behaviours, for example by including preferences for increasing social welfare.
In this paper, we will refer to the specific case of reward shaping involving only the extrinsic rewards of other agents as \text{intrinsic motivation}, because it represents the agents having social motivations beyond their own selfish reward.
% One class of reward shaping methods uses an intrinsic reward proportional to the extrinsic rewards of other agents.
% Before we start introduce the algorithms, we want to mention that, existing literature frequently employ \textit{intrinsic motivation} design to induce useful cooperative behaviours.
Specifically, let $r^{\text{ex},i}$ denote extrinsic environment reward for agent $i$ and   $r^{\text{in}}$ denote the intrinsic motivation, agent $i$'s immediate reward is modified as follow,

\begin{equation}
\label{eq:intrinsic-motivation}
    r^i := r^{\text{ex},i} + r^{\text{in},i}.
\end{equation}

\paragraph{Altruism}
A particularly common form of intrinsic motivation is called \text{Altruism}.
While in colloquial use, altruism begets notions of self-sacrifice, in this field it refers to an agent caring about the wellbeing of others, but not necessarily at their own expense.
For example, the model of altruism proposed by \cite{chen2008altruism} is as follows.
An agent receives a reward that is composed of a proportion of their extrinsic game reward, $p_i(s)$, and a proportion of the mean group social welfare, $SW(\bm{r})$.
A parameter $\beta \in [0,1]$ controls the tradeoff between the two.

\begin{equation}
    % \nonumber
    \forall i \in N, \quad r^i(\beta) = (1 - \beta) r^{\text{ex},i} + \frac{\beta}{n} SW(\bm{r}^{\text{ex}}).
\end{equation}

% For every $\beta \in[0,1], G(\beta):=\left(N,\left\{S_i\right\}_{i \in N},\left\{r_i^\beta\right\}_{i \in N}\right)$ with
% $$
% r_i^\beta(s)=(1-\beta) p_i(s)+\frac{\beta}{n} S W(s) \quad \forall i \in N .
% $$

If agents care about the collective return, this directly addresses the core problem of a social dilemma - that there are opportunities to profit at the expense of the group.
Now, though an agent maintains their own interests, if they take actions that harm the collective, this too will lower their own reward.
Additional models of altruism are discussed by \cite{apt2014selfishness}.

% \begin{equation}
%     r_i+\alpha \text{SW}.
% \end{equation} 
% Model A \citep{elias2010socially}: 
% For every $\alpha \geq 0, G(\alpha):=\left(N,\left\{S_i\right\}_{i \in N},\left\{r_i^\alpha\right\}_{i \in N}\right)$ with
% $$
% r_i^\alpha(s)=p_i(s)+\alpha S W(s) \quad \forall i \in N
% $$
% Model C \citep{chen2011robust}: For every $\gamma \in[0,1], G(\gamma):=\left(N,\left\{S_i\right\}_{i \in N},\left\{r_i^\gamma\right\}_{i \in N}\right)$ with
% $$
% r_i^\gamma(s)=(1-\gamma) p_i(s)+\gamma S W(s) \quad \forall i \in N .
% $$
% Model D \citep{caragiannis2010impact}: For every $\delta \in[0,1], G(\delta):=\left(N,\left\{S_i\right\}_{i \in N},\left\{r_i^\delta\right\}_{i \in N}\right)$ with
% $$
% r_i^\gamma(s)=(1-\delta) p_i(s)+\delta\left(S W(s)-p_i(s)\right) \quad \forall i \in N
% $$

% The selfishness level notion for Model A extends to Models B, C and D. 
% Unlike the price of stability and anarchy, this measure remains consistent across payoff function modifications.
% By examining various strategic games, the research highlights the balance between individual interests and societal impact. Some games show a finite selfishness level, while others can have infinite levels. The paper provides specific bounds for particular games and delves into the factors affecting player cooperation within those games. 
% Social optimality can be proved.

\paragraph{Game extensions}
Another popular method is to extend the game to include additional mechanisms, which agents can utilise with an expanded action space.
An example of this is contracting~\citep{hughes20__learning_to_resolve_alliance_dilemmas_in_manyplayer_zero-sum_games,christoffersen2023get}. In one version of this idea, at each timestep, a pair of agents can propose a joint-action.
If both agents propose the same joint-action, then they must take their corresponding action, otherwise both agents choose their actions as normal.
Other extensions include methods where agents explicitly affect the rewards of other agents through dedicated means, such as by gifting~\citep{lupu2020gifting} or exchanging reward~\citep{willis2023resolving}.

\paragraph{Models of the emergence of cooperation}
Much of the literature in this area has been concerned with the question of how cooperation of self-interested agents may come about and remain stable, despite the ever-present opportunities for conflict, overconsumption, free-riding, and defection that threaten it. This line of work is concerned with modeling humans, and as such is intended to connect with long-established theories in economics and other social sciences \citep{hertz2023beyond} which have been interested in self-interested conceptions of agent motivation (in part as a result of adherence to the methodological individualism used in these fields \citep{heath2020methodological}). Moreover, it is preferable to exhibit a simpler account of cooperation over a more a complex one on grounds of parsimony. A common refrain in this area is to say that one is ``\text{searching for the minimal set of maximally general priors}''.

\begin{table}[t!]
\centering
\caption{Summary of representative algorithms for resolving SSDs}
\label{tab:algs-mixed-motive}
\vspace{3mm}
\begin{tabular}[t]{p{2cm}|p{5cm} |l}
\hline
Types & Description & Algorithms \\
\hline
Other-regarding preferences  & Agents either intrinsically care about the welfare of others, or they modify the extrinsic game rewards of other agents & 
\begin{tabular}[t]{@{}l@{}}
Prosocial \citep{peysakhovich2018prosocial}\\
Inequity aversion \citep{hughes2018inequity}\\
Evolving motivations \citep{wang2019evolving}\\
PED-DQN \citep{hostallero2020inducing}\\
SVO \citep{mckee2020social}\\
Gifting \citep{lupu2020gifting}\\
Gifting for prosociality \citep{wang21__emergent_prosociality_in_multiagent_games_through_gifting}\\
Relational networks \citep{haeri2022reward}\\
D3C \citep{gemp2022d3c}\\
Auto-aligning incentives \citep{kwon23__autoaligning_multiagent_incentives_with_global_objectives}\\
\end{tabular}\\
\hline
Other-influence & Agents consider how their actions will impact the future behaviour of their fixed co-players &
\begin{tabular}[t]{@{}l@{}}
Hierarchical social agency \citep{kleiman2016coordinate}\\
CCC \citep{peysakhovich2018towards_AI}\\
LOLA \citep{foerster2018learning}\\
Imitation \citep{eccles2019learning}\\

Cooperation degree \citep{wang2019achieving}\\

SOS  \citep{letcher19__stable_opponent_shaping_in_differentiable_games}\\
Social influence \citep{jaques2019social}\\
LIO \citep{yang2020learning}\\
Cooperative learning \citep{jacq20__foolproof_cooperative_learning}\\
M-FOS  \citep{lu22__modelfree_opponent_shaping}\\

\end{tabular} \\
\hline
Reputation and Norms & Agents assess whether their co-players comply with \text{social rules} and modify their behaviour accordingly &  
\begin{tabular}[t]{@{}l@{}}
Competitive altruism \citep{mckee2021deep}\\
Reputation dynamics \citep{anastassacos21__cooperation_and_reputation_dynamics_with_reinforcement_learning}\\
CNM \citep{vinitsky2023learning} \\
\end{tabular} \\  
\hline
Contracts &   Agents are able to commit to taking joint-actions or promises of future rewards  &
\begin{tabular}[t]{@{}l@{}}
Contracts \citep{hughes20__learning_to_resolve_alliance_dilemmas_in_manyplayer_zero-sum_games}\\
RUSP \citep{baker20__emergent_reciprocity_and_team_formation_from_randomized_uncertain_social_preferences}\\
Contracts with payments \citep{christoffersen2023get}\\
Reward exchange \citep{willis2023resolving}\\
Reward shares \citep{schmid23__learning_to_participate_through_trading_of_reward_shares}\\

\end{tabular} 
\\ \hline
\end{tabular}
\end{table}
\subsection{Solution Approaches}

We now review some popular classes of approaches for tackling SSD problems.
Table~\ref{tab:algs-mixed-motive} summarises the related algorithms, where we have classified them based upon the mechanism that they use.
% Most of these involve agents simply augmenting their own reward using some observation of others' behaviours.
% Based on the general principles that guide the cooperation, the existing solutions can be classified into four categories, Other-regarding preferences, reciprocity, and norms.

\paragraph{Other-regarding preferences}
% \yali{Add reference to explain. }
% \yali{Altruism}
% One class of reward shaping methods uses an intrinsic reward proportional to the extrinsic rewards of other agents.
When an intrinsic reward is proportional to the mean collective reward \citep{peysakhovich2018prosocial}, it is regarded as a \text{prosocial reward}, and it is equivalent to altruism.
% , and has been shown to help reinforcement learning agents converge to a better equilibrium in a social dilemma called Stag Hunt (Table \ref{table:sd_c}).
This same collective reward was used by \cite{hostallero2020inducing}, though they constrained the intrinsic reward to use only those of other agents within an agent's observation.
The authors argue that by limiting the intrinsic motivation to include the reward of \text{local} agents, this can help with the credit assignment problem (introduced in Section~\ref{sec:credit}).

\citet{hughes2018inequity} split the typical altruism reward into two parts: \text{advantageous} inequity or \text{guilt} when an agent outperforms the mean, and \text{disadvantageous} inequity or \text{envy} when an agent underperforms the means.
The intrinsic motivation for agent $i$'s is given as
\begin{equation}\label{eq:inequity-aversion}
    r^{\text{in}, i} :=- \frac{\alpha}{N-1} \sum_{j \neq i}{\max(e_t^j(s_t^j, a_t^j) - e_t^i(s_t^i, a_t^i), 0)} - \frac{\beta}{N-1} \sum_{j \neq i}{\max(e_t^i(s_t^i, a_t^i) - e_t^j(s_t^j, a_t^j), 0)}.
    % r^{\text{in}, i} :=- \frac{\alpha}{N-1} \sum_{j \neq i}{\max(r_t^{\text{ex},j}(s_t^j, a_t^j) - r_t^{\text{ex},i}(s_t^i, a_t^i), 0)} - \frac{\beta}{N-1} \sum_{j \neq i}{\max(r_t^{\text{ex},i}(s_t^i, a_t^i) - r_t^{\text{ex},j}(s_t^j, a_t^j), 0)}
\end{equation}
Therefore, the immediate reward for agent $i$ is given as  $r^i := r^{\text{ex},i} + r^{\text{in}, i}$, where  $r^{\text{ex},i}$ represents the extrinsic reward. 
% \yali{stop here}
% Here, $r^i(s_t^i, a_t^i)$ represents the extrinsic reward 
$e_t^j(s_t^j, a_t^j)$ for agents $j = 1,...,N$ are their extrinsic rewards smoothed in a manner analogous to eligibility traces \citep{sutton2018reinforcement}:
% For sequential games, 
\begin{equation}
    e_t^j(s_t^j, a_t^j) := \gamma \lambda e_{t-1}^j(s_{t-1}^j, a_{t-1}^j) + r^{\text{ex},j}(s_t^j, a_t^j),
\end{equation}
for discount factor $\gamma$ and hyperparameter $\lambda$.
$\alpha$ and $\beta$ are hyperparameters which respectively control the importance of envy and guilt.
The authors experimentally found that in different social dilemmas, different proportions of envy and guilt performed best.
While \cite{hughes2018inequity} determined these hyperparameters with a grid search, \citet{wang2019evolving} determined them  using a technique called Population Based Training \citep{jaderberg2017population} which they interpreted as a model of evolution operating on the reward function. 
 In the variant where cooperation emerged it was specifically a model of multi-level (group) selection \citep{duenez2023social}, this was an expected result since it has been known since Darwin that individual fitness maximization does not on-its-own lead to the evolution of altruistic traits \citep{nowak2006evo}.

\begin{wrapfigure}{r}{0.5\textwidth}
% \begin{figure}[th]
\centering
\includegraphics[width=8.5cm]{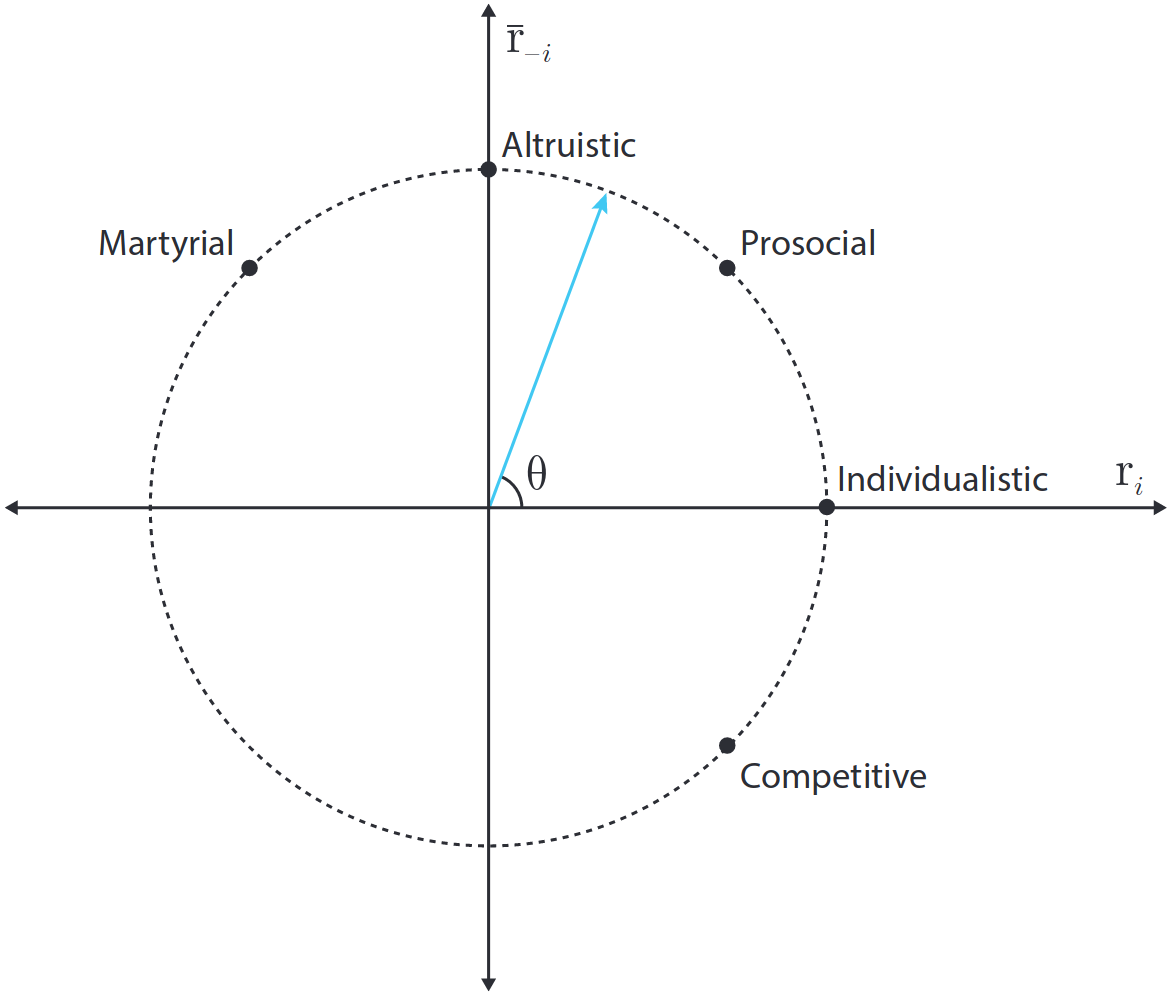}
\caption{\label{fig:svo}An agent's Social Value Orientation can be expressed by the parameter $\theta$, which specifies the ideal relationship between an agent's own reward and that of the other agents}
% \end{figure}
\end{wrapfigure}
A general method for constructing different attitudes or preferences an agent may have for the relationship between their own reward and the mean group reward is is Social Value Orientation (SVO) \citep{mckee2020social}, which takes inspiration from interdependence theory.
The social values of an agent can be specified using a point on a circle centred at the origin where the $x$-axis is the agent's reward and the $y-axis$ is the arithmetic mean $\Bar{r}_{-i}$ of all other agents' rewards.
Given reward vector $\textbf{r}$, the reward angle $\theta^i(\textbf{r})$ for agent $i$ satisfies
\begin{equation}
  \tan{\theta^i(\textbf{r})} = \frac{\Bar{r}^{-i}}{r^i}  
\end{equation}
and an agent's preferred reward angle $\theta_{SVO}^i$ is its SVO.
The reward is then augmented with an intrinsic reward penalising deviation of the reward angle (given observed rewards) from the SVO:
\begin{equation}
  R^i(s_t^i, a_t^i) := r^i(s_t^i, a_t^i) - w |\theta_{SVO}^i - \theta^i(\textbf{r})|.  
\end{equation}
This is depicted in Figure~\ref{fig:svo}.
The authors demonstrate the positive effects of heterogeneity on achieving complex behavioural variation in a manner consistent with interdependence theory.
The heterogeneous agents ultimately outperform homogeneous populations on the SSDs considered.
Somewhat analogously, the final method \citep{haeri2022reward} uses pre-defined social structures, represented with a graph, and qualitatively assesses the performance of these relationships in the presence of agent imbalance.

% The next method we will consider leverages population-based training (PBT) \citep{jaderberg2017population} to evolve the intrinsic reward.
% More specifically, the intrinsic reward is modelled using a 2-layer neural network and can be formulated as

% \[u^i(f^i | W, v, b) := v^T \omega (W^T f^i + b)\]

% where the parameters $W$, $v$ and $b$ of the network are evolved using a fitness function $F^i$, $v$ is a hyperparameter vector and $f^i$ is a feature vector consisting of either smoothed observed rewards (retrospective) or value estimates (prospective) of all agents sampled for play in the current episode.
% The fitness function $F^i$ is a moving average of total episode return for agent $i$.
% Since the (positive) intrinsic reward is added to the extrinsic reward, this encourages training to improve other agents' rewards as well as one's own.
% Moreover the parameter evolution ensures this altruism is trained in a way which optimises for the given agent's own long-term reward.
% Agents thus learn inductive biases for cooperation in a model-free way.

Rather than having intrinsic motivations, which relies upon agents adopting preferences beyond their own self-interest, by transferring reward an agent is able to modify the extrinsic game reward of their co-players.
If this modification leads to a more cooperative approach in the recipients, the agent transferring rewards can experience a net benefit.
The most simple method is \text{gifting} \citep{lupu2020gifting,wang2021towards}, where agents are able as part of their actions to give rewards directly to other agents, in a zero-sum manner.
This mechanism allows agents to reward their peers for cooperation.
\citet{gemp2022d3c} introduce D3C which is also used by \citet{kwon23__autoaligning_multiagent_incentives_with_global_objectives}.
This algorithm takes a collective approach, and finds an optimal mixture of \text{reward transfer} or \text{loss sharing} to improve the worst-case performance of the group.

\paragraph{Other-influence}
An agent can strategically encourage cooperative behaviour by mirroring the prosocial behaviour of others.
In this way, an agent remains robust to exploitation while offering to engage in reciprocal cooperation with an co-player, as inspired by the game theory strategy for iterated normal-form social dilemmas called \text{Tit-For-Tat} \citep{axelrod80__effective_choice_in_the_prisoners_dilemma}.
Whereas altruism incentivises an agent to take prosocial actions regardless of the behaviour of other agents, here prosocial actions are dependent on those of other agents.
A common approach is to train two policies, a cooperative, prosocial policy, and a self-interested policy.
If the co-player is deemed to be behaving pro-socially, typically as assessed by a classifier \citep{wang2019achieving, kleiman2016coordinate} or by analysis of one's own rewards \citep{peysakhovich2018towards_AI}, then the agent uses the prosocial policy, otherwise it uses the self-interested policy.
In the case of \citet{wang2019achieving}, a full range of cooperative behaviours is achieved by synthesising the two policies to match the degree of cooperativeness exhibited by the co-player.

In \citet{eccles2019learning}, the authors use two types of agents; innovators and imitators.
Innovators are standard learners whose rewards are purely extrinsic (selfish).
Imitators will include an intrinsic reward term which seeks to minimise the difference in \text{niceness} between themselves and an innovator, thus mimicking the cooperativeness of the innovator.
Specifically, suppose we have an imitator ($\text{im}$) and an innovator ($\text{in}$) and associated trajectories $T^{\text{im}}$ and $T^{\text{in}}$. The imitator's intrinsic reward is defined as
\begin{equation}
r^{\text{im}}(t) := -(N^{\text{im}}(T^{\text{im}}) - N^{\text{in}}(T^{\text{in}}))^2,
\end{equation}
where the \text{niceness function} $N^i(T^i)$ for agent $i \in \{\text{im}, \text{in}\}$ measures the effect on an agent's reward of the other agent's actions through a trajectory. Formally, define first the niceness of an innovator action for the imitator as
\begin{equation}
    n_{\pi_{\text{in}}}^{\text{im}}(s_t^{\text{in}}, a_t^{\text{in}}) := Q_{\pi_{\text{in}}}^{\text{im}}(s_t^{\text{in}}, a_t^{\text{in}}) - V_{\pi_{\text{in}}}^{\text{im}}(s_t^{\text{in}}),
\end{equation}
where $V_{\pi_{\text{in}}}^{\text{im}}$ and $Q_{\pi_{\text{in}}}^{\text{im}}$ are neural network models for the value and the action-value functions respectively.
The niceness function of an imitator action is defined equivalently.

% \[n_{\pi_{im}}^{in}(s_t^{im}, a_t^{im}) := Q_{\pi_{im}}^{in}(s_t^{im}, a_t^{im}) - V_{\pi_{im}}^{in}(s_t^{im})\]

These functions quantify the effect of the other agent's action on one's own reward. Note that the action-value functions are estimating the discounted return to the agent from time $t$ given the other agent's states and actions.
The niceness function of a trajectory for agent $i \in \{\text{im}, \text{in}\}$ can then be defined w.r.t. agent $j \in \{\text{im}, \text{in}\}\\i$ as the discounted sum of action niceness values
\begin{equation}
N^{i}(T^i) := \sum_{k=1}^{t}{\gamma^{t-k}}n_{\pi_{i}}^{j}(s_k^i, a_k^i),
\end{equation}
where $t$ is the length of the trajectory $T^i$. This niceness function can be used to calculate the intrinsic reward, which is normalised before being added to the extrinsic reward.
% This method shows how agents can learn to cooperate through reciprocal behaviour and accordingly the agents develop tit-for-tat like strategies in the experiments.

An independent algorithm that treats other agents as part of the environment is called a \text{naive learner}.
Because the environment is assumed to be static, naive learners do not appreciate that their co-players may change their policies, and so not do take into account how changes in their own policy can cause responses in their co-players.
The following methods take into account the learning of their co-players, and are called \text{co-player shaping}. 
% and can be applied to zero-sum games where 'opponent' is mostly used.
% \textcolor{red}{JZL:  It doesn't really make sense to call it `opponent' shaping in the mixed-motive case. That terminology is really just for zero-sum. Here we should call it something else like 'co-player shaping' perhaps.}
% \rich{I disagree here, LOLA and SOS use Prisoner's dilemma, and MFOS uses the coins game. I think it is reasonable to stick to the current terms, rather than introduce a new one.}
Co-player shaping is a rare departure from the usual intrinsic reward mechanism, because the other agents' policy parameters are used directly in the value function optimisation stage.
Unlike \text{reward transfer} methods, where an agent will directly modify the extrinsic rewards of their co-player and leave them to learn how to optimise it, co-player shaping directly considers the policy updates their co-player will make.
The during training, an agent using this method attempts to ensure that their co-player will learn to take actions that benefit the agent using  co-player shaping.

The first method to introduce this technique was LOLA \citep{foerster2018learning}, which assumed it was facing a naive learner.
Instead of optimising the standard value function $V(\theta_1, \theta_2)$, where $\theta_1$ and $\theta_2$ are the policy parameters of the current agent $1$ and those of the other agent respectively, this method optimises $V(\theta_1, \theta_2 + \Delta \theta_2)$, where $\Delta \theta_2$ is one naive update of agent $2$'s parameters.
The other agent's parameters are inferred from its state-action trajectories through a form of co-player modelling similar to behaviour cloning.
This method is able to achieve cooperation via the following approach: when an co-player takes the action that you want them to, reciprocate so that they receive higher reward and are more likely to take this action next time.
Otherwise, punish them to dissuade other actions.

\citet{letcher19__stable_opponent_shaping_in_differentiable_games} and \citet{lu22__modelfree_opponent_shaping} improved upon  LOLA, increasing its robustness and relaxing requirements to have a differentiable model of their co-player.
\citet{jacq20__foolproof_cooperative_learning} extended these concepts to include arbitrary learning algorithms for $n$-player games, and developed an approach whereby a group of agents can retaliate against any selfish behaviour.

A similar method, Learning to Incentivise Others (LIO) \citep{yang2020learning} extends co-player  shaping to include reward transfers to the co-player, which allows an agent to shape the behaviour of their co-player directly, without needing to influence the co-player’s 's extrinsic rewards through the consequences of their own actions alone.
Here, a reward giver agent $i$ learns an incentive function $r^{\eta^i}: O \times A^{-i} \xrightarrow{} \mathbb{R}^{N-1}$ where $O$ is its own observation space, $A^{-i}$ is the joint action space of all other agents and $\eta^i$ is a parameter vector.
The incentive function determines the rewards given by $i$ to the other agents based on its observation of the state and their actions.
Consequently, a recipient agent $j$ has a new reward $r^j$.
Let $r_{\eta^i}^j$ be the reward given by $i$ to $j$. Then $j$'s new reward is

\begin{equation}
r^j(s_t, \boldsymbol{a_t}, \eta^{-j}) := r^{\text{ex},j}(s_t, \boldsymbol{a_t}) + \sum_{i \neq j}{r_{\eta^i}^j(o_t^i, \bm{a_t}^{-i})}.
\end{equation}
While $i$ optimises its incentive function $r_{\eta^i}$ to maximise a separate objective, $r^j$ is then optimised as usual with policy gradients. The latter consists of a positive term representing $i$'s expected extrinsic reward due to the reward recipients' new actions as well as a negative term representing the discounted sum of rewards given (i.e. a cost for reward giving). This approach facilitates gifting with gifted rewards determined by a dedicated selfish objective, thereby encouraging cooperation to emerge from self-interest. Notice that the policy updates require knowledge of other agents' parameters. The authors relax this requirement with co-player modelling using other agents' observed states and actions.

Another method uses the idea that agents may seek to maximize their social influence \citep{jaques2019social}, i.e. the extent to which their action changes the decisions of others. Mathematically, the full reward is $r^i := \alpha r^{\text{ex},i} + \beta r^{\text{in},i}$ where $r^{\text{in},i}$ is calculated as the sum over other agents $j \neq i$ of KL-divergences between $j$'s policy conditioned on $i$'s action and $j$'s marginal policy (independent of $i$'s action):
\begin{equation}
r^i := \sum_{j \neq i}{D_{KL}[p(a_t^j | a_t^i, s_t^j) || p(a_t^j | s_t^j)]}.
\end{equation}
The quantity $p(a_t^j | s_t^j)$ is estimated as $\sum_{\Tilde{a}_t^i}{p(a_t^j | \Tilde{a}_t^i, s_t^j)p(\Tilde{a}_t^i | s_t^j)}$, where $\Tilde{a}_t^i$ are counterfactual samples of $k$'s action. Agent $i$'s intrinsic reward can be seen as a measure of $i$'s social influence and adding this to the reward encourages choosing more socially influential actions, thus leading to increased cooperation. In a second level of their method, the authors introduce an explicit communication channel. Namely, each agent outputs an extra vector $m_t^i$ representing a message. All the agents' messages are concatenated and then provided as input to each agent in the next step. The intrinsic reward in this case is similar to the above but instead of conditioning agent $j$'s distribution on $i$'s action $a_t^i$, it is conditioned on $i$'s message $m_{t-1}^i$ from the previous step. This has a similar effect of representing social influence but this time assuming that influence travels through the dedicated communication channel. Empirical results show improved coordination, communication and collective return due to social influence.

\paragraph{Reputation and norms}
A norm is a collective pattern of behaviour supported by a shared pattern of sanctioning. 
Some work has taken inspiration from how humans use their own reputation within the group as motivation to cooperate.
One such method is \citet{mckee2021deep}, where the reward is as before a sum of extrinsic reward and an intrinsic reward, calculated as
\begin{equation}
r^{\text{in},i} := -\alpha \max(\Bar{c} - c^{\text{self}}, 0) - \beta \max(c^{\text{self}} - \Bar{c}, 0),
\end{equation}
where $c^{\text{self}}$ is a measure of one's own contribution level and $\Bar{c}$ is the observed or estimated average group contribution level.
This equation is analogous to inequity aversion described in Eq. \eqref{eq:inequity-aversion} (but with contribution levels instead of smoothed rewards) and accordingly $\alpha$ and $\beta$ are scalar parameters.
The terms in this intrinsic reward penalise large discrepancies between one's own contribution level and that of the group as a whole.
The authors build a computational model of human behaviour and show that humans can effectively cooperate on the Cleanup task when other players are identifiable and their reputations can be tracked.
However, they fail to cooperate under conditions of anonymity.
The MARL agents also demonstrate these behaviours with regard to identifiability and anonymity.
\citet{anastassacos21__cooperation_and_reputation_dynamics_with_reinforcement_learning} assess which reputation norms lead to the emergence of cooperation in a population of reinforcement learning agents playing matrix game social dilemmas.

In the Classifier Norm Model (CNM) \citep{vinitsky2023learning}, agents learn social norms through \text{public sanctioning}. This is inspired by human societies, where people adjust their behaviour based on their internal understanding of the behaviours to which society as a whole gives its approval or disapproval. In this model, agents have opportunities to sanction one another using a grounded in-game mechanism, whereby they may zap each other with a \text{punishment beam} that causes negative reward in any player hit by it. Agent A has an opportunity to sanction agent B whenever agent B is in range of their zapper. Whenever zapping events occur it is assumed that everyone gossips about them, so there is public knowledge of sanctioning opportunities and whether or not they led to a zap, which is interpreted as disapproval of the agent who was zapped's recent behaviour, or did not lead to a zap, in which case it would be interpreted as approval of the not-zapped agent's recent behaviour. Each agent learns a personal  representation of what behaviour is acceptable to its group. This takes the form of a classifier of whether or not a given behaviour is likely to provoke approval or disapproval. The classifier is trained with all the group's public sanctioning data. Finally, agents are assumed to have an intrinsic motivation to align their own sanctioning behaviour (zapping) with the predictions of the classifier they learned (i.e.~with their personal representation of social norm). They are motivated to zap the in the same context that others in their group would also zap. Therefore, this system exhibits a runaway bandwagon effect where most agents then end up supporting a particular norm, i.e.~sanctioning the same behaviour as one another and complying with the norm by refraining from emitting the sanctioned behaviour. This is a model where the content of a social norm proscribing a behaviour emerges from the dynamics of multi-agent learning. 

The CNM model is in line with a decentralised vision for mixed-motive agents. For example, the agents could be seen like models of autonomous vehicles learning the local norms of a new town by observing sanctioning events such as other cars honking their horns at one another. Humans may also participate in this using the same interface for sanctioning as the agents (e.g.~humans already understand the social meaning of honking a car horn).

\paragraph{Contracts}
A separate paradigm for encouraging cooperation involves allowing agents to more directly affect other agents' rewards through dedicated structures rather than indirectly through the environment.
For example, {contracts} \citep{hughes20__learning_to_resolve_alliance_dilemmas_in_manyplayer_zero-sum_games} can be used for this purpose.
A contract is a joint state-action dependent vector of rewards which can be proposed and accepted by agents.
The authors formulate an augmented game where agents' actions are augmented with the ability to propose contracts, and if both players propose the same joint-action, it becomes binding for the next timestep.

\citet{christoffersen2023get} extended these contracts to include a possible side payment. This approach gives agents the ability to share rewards with other accepting agents, given that the accepting agents follow action trajectories satisfying given specifications.
This can be thought of as committing to reward transfer conditional on behaviour.
For example, a contract could be interpreted as ``I'll pay you to clean the river (so that I can pick apples)''.

Rather than specifying contracts on an action-level, which can be burdensome, agents could alternatively enter an agreement that covers the whole episode.
\citet{schmid23__learning_to_participate_through_trading_of_reward_shares} allow agents to buy and sell stakes in the future rewards of their co-players, in an analogous manner to buying shares in a company, while \citet{willis2023resolving} find the minimum proportion of reward that agents must commit to exchanging between themselves over an episode to learn cooperative policies in a sequential social dilemma.
In \citet{willis2023JAAMAS}, the authors generalise their method to allow unequal transfers between agents and find transfer arrangements that resolve matrix game dilemmas while retaining the greatest possible proportion of their own extrinisic reward or self-interest.
\citet{baker20__emergent_reciprocity_and_team_formation_from_randomized_uncertain_social_preferences} uses a sampling approach and randomises the transfers (which can be thought of as relationships) between the agents.
Furthermore, the agents are uncertain about these relationships.
Their experiments suggest that both of these features help the learning algorithms to develop reciprocal behaviours.

\section{Evaluation} \label{sec:eval}
% As there are many benchmark tasks, including notably the Atari game suite ALE \citep{Bellemare_2013}, that has served the development of single-agent reinforcement learning algorithms, the MARL community has developed its own range of benchmarks as well as having a larger range of possible evaluation metrics. 
% We will below review a number of the most important benchmark suites, but it is also worth mentioning the work that has been developing algorithms for competing with the best humans in popular and very challenging multi-player games including Go, Chess and Shogi \citep{schrittwieser2019mastering}, Poker \citep{mel2017poker,brown2019poker}, Starcraft II \citep{vinyals2019grandmaster}, Strategyo \citep{perolat2022stratego} and diplomacy \citep{bakhtin2022diplomacy}.

The evaluation of MAL algorithms can be challenging, it has been noticed that there is a substantial variability in reported results even for the same algorithm on the same task \citep{gorsane2022towards}.
The community calls for standardised evaluation protocols, focusing on default parameters such as training time, standardised uncertainty quantification and more complete reporting on failure cases. Below, we provide a review of representative tasks and evaluation metrics.
% In both single-agent and multi-agent RL benchmarking it has been noticed that there is a substantial variability in reported results even for the same algorithm on the same task \citep{gorsane2022towards}, and a standardised evaluation protocol was proposed focused around default parameters such as training time, standardised uncertainty quantification and more complete reporting.

\subsection{Environments}
% Along with the \yali{stop here}
Aligning with the methodology described in Section \ref{sec:team-based} and \ref{sec:mixed-motive}, we describe the benchmarks for fully cooperative team-based tasks and mixed-motive tasks respectively. Below are open-sourced simulators and benchmarks for fully cooperative tasks.
\begin{itemize}
    \item[$-$] {\bf StarCraft Multi-agent Challenge (SMAC):} StarCraft II \citep{vinyals2017starcraft} is a popular testbed for MARL algorithms, presenting various battle scenarios where agents correspond to {units} (i.e. combatants) who must cooperate to defeat computer controlled opponents. SMAC has become a standard evaluation suite for MARL, much like Atari in the single-agent case.
% \yali{Exaamples likx xxx algirhtms used this ..}
\item[$-$] {\bf Overcooked:} Overcooked \citep{carroll2020utility} is a video game environment where agents must cooperate to prepare and serve onion soup. They must split cooking/serving responsibilities among each other and learn to work with varied teammates.
% \yali{Exaamples likx xxx algirhtms used this ..}

\item[$-$]
{\bf MuJoCo Soccer Environment:} The MuJoCo Soccer environment \citep{liu2019emergent} is a simple 2v2 team game where agents can move in 3D space to kick a ball into the opponent team's goal. This type of environment allows investigation of emergent cooperative behaviours such as ball chasing.
% \yali{Exaamples likx xxx algirhtms used this ..}

\item[$-$]
{\bf Google Research Football (GRF):} A much more realistic football game is Google Research Football \citep{kurach2020google}, which resembles popular football games available on games consoles. Here, all the standard rules of the game apply such as corner kicks, fouls, cards, kick-off, offside, etc. In addition, the physical representation of the players is highly realistic. This environment allows for a much more complex range of learning behaviours to be studied, alongside customising the difficulty level.
% \yali{Exaamples likx xxx algirhtms used this ..}
\end{itemize}

Recently, a surging interest is seen in enhancing cooperation in mixed-motive tasks. Some representative tasks are described below.
\begin{itemize}
    \item[$-$] {\bf Clean-up:} Clean-up \citep{hughes2018inequity} was inspired by social dilemmas of public good provision. Agents gain reward by harvesting apples available in an orchard. There is a river which feeds the apple orchard. However, pollution is being dumped into the river from outside the environment so it fills up with pollution with a constant probability over time. As the proportion of the river filled with pollution increases, the growth rate of apples monotonically decreases. No apples grow at all once pollution levels exceed a threshold. Individuals can spend time cleaning the river to remove its pollution, an extended course of action that is analogous to making a contribution to the public good of size proportional to the amount of time they clean and their skill in doing so. This is a sequential social dilemma due to the temptation of neglecting the river cleaning to instead free-ride by collecting apples in the orchard, a course of action which leads to ruin if all elect it simultaneously.
    
    \item[$-$] {\bf Commons Harvest:} Commons Harvest was inspired by common-pool resource appropriation scenarios (see \citet{janssen2010lab}). The MAL environment was introduced in \citet{perolat2017multi} under the name Commons. Later it was renamed to Harvest \citep{hughes2018inequity}. Subsequent work used both names interchangeably. More recently it has usually been called Commons Harvest to try reduce the naming confusion \citep{leibo2021scalable, agapiou2022melting}. In Commons Harvest agents must navigate a 2D world to collect apples. The apple spawn rate in each location depends positively on the number of nearby apples, so that they grow in groups. If there are no apples in a local area then the probability of new growth in that area is zero. This is a sequential social dilemma because there is conflict between the short-term reward of collecting all apples in a particular area vs the long-term cost of apples never growing back in the area. Most of the mixed-motive methods we have discussed were evaluated on both Clean-up and Commons Harvest.

    \item[$-$] 
{\bf Coin Game:}
Another environment that is commonly used by the discussed methods in Coin Game \citep{lerer2017maintaining}, where agents navigate a gridworld to collect randomly placed coins, each of a randomly assigned colour. Agents obtain a reward of +1 for collecting any coin but if it is of the other agent's colour, the latter receives a reward of -2. If both agents succomb to this temptation, their expected reward is 0. Agents must therefore learn to sacrifice and coordinate based on colours in order to maintain long-term rewards.

    \item[$-$] {\bf Level-Based Foraging (LBF):} In the LBF tasks \citep{christianos2020shared, papoudakis2021benchmarking}, agents navigate a grid-like domain where their goal is to gather items. These items and agents have distinct levels assigned to them. To successfully gather an item, the combined levels of cooperating agents must meet or surpass the item's level. Successfully acquiring an item grants agents a reward corresponding to the item's level. Full visibility of the environment is the default setting for agents, but a variation limits their vision to a nearby 5x5 square.
    
    \item[$-$] 
    {\bf Melting Pot:} Melting Pot \citep{leibo2021scalable, agapiou2022melting} is an evaluation suite tailored  to assess the capability of MARL algorithms to interact and adapt when faced with unfamiliar agents within familiar games that are here called substrates. A substrate together with a population of agents to face is called a scenario. The suite encompasses a breadth of social dynamics, such as cooperation, competition, deception, and trust in a range of over 50 substrates (for the expanded Meltingpot 2.0) and over 250 scenarios. After training agents on specific multi-agent games, the real test comes from evaluating their adaptability with new agents in the same game contexts. The unique aspect of Melting Pot is its scalability: by simply introducing a new set of opponent agents, an entirely new evaluation challenge is created, without altering the inherent game.
    Melting Pot brings together many pre-existing environments, including several of those mentioned above: Overcooked (which it calls Collaborative Cooking), Clean-up, Commons Harvest, and Coin Game.

\end{itemize}

\subsection{Metrics}
\label{sec:metrics}

% In multi-agent experiments, the evaluation landscape is diverse, driven by the nature of agent interactions and the overarching objectives of the experiment. The metrics typically center around individual and collective performance:
% \begin{itemize}
%     \item \textbf{Individual Returns} quantifies each agent's individual performance.
%     \item \textbf{Total Utility for the Population} aggregates the returns of all agents. In fully cooperative scenarios, this metric is (or is proportional to) the team return.
% \end{itemize}

% Besides these obvious metrics, the richness of the multi-agent setting introduces a myriad of metrics depending on the context:

% \begin{itemize}
%     \item \textbf{Ranking:} In competitive scenarios where agents or teams oppose each other, ranking based on performance is typically involved.
    
%     \item \textbf{Social metrics:} In experiments simulating economic systems or social dilemmas we are also interested in metrics related to social welfare, equality and peace.
  
% \end{itemize}

We will now outline the evaluation metrics used in the literature for the methods we have discussed. The choice of metric depends on the principles being evaluated. In the team game case, there is a common reward for all agents at each step and all incentives are therefore the same. This makes metric choice easy as there is only one option, namely the value of the common reward achieved during an episode (or equivalently the common win/success rate). Accordingly, all team game methods discussed use the common reward or win rate as their evaluation metric. 
% One slight exception here is MADDPG, where it is technically possible for agents to have distinct rewards but in this case the reward is averaged over agents.
One slight variation here is that agents have distinct rewards but still the collective reward is considered for the team performance. 

In the mixed-motive setting, the situation is much more complex  as here there is an inherent conflict between agents' individual incentives and those of the group. Which incentives do we prefer the system to optimise, those of the individual or the group? Indeed how do we even measure group incentives? The methods we have discussed for this setting all agree on the importance of \text{collective return} or \text{utilitarian} metric, which is the total episodic reward received over all agents:

\begin{equation}
R_C = \sum_{i=1}^{N} R^i.
\end{equation}

Note that in some cases this is substituted by the average return over agents but the concept here is identical. Most works additionally employ some \text{social} metrics aimed at quantifying the level of cooperativeness achieved. We find that, although a selected few metrics are used by multiple works, there is in general little consensus on how to measure cooperativeness and the metrics tend to be heavily task-specific or subjective. In the following we will go through the social metrics used in the mixed-motive literature. 

\textbf{Sustainability} Capturing the notion of sacrificing selfish immediate rewards in order to maintain long-term rewards for the group, sustainability measures the average time at which rewards are achieved:
\begin{equation}
S = \mathbb{E}[\frac{1}{N}\sum_{i=1}^{N}{t^i}],
\end{equation}
where $t^i = \mathbb{E}[t | r_t^i > 0]$. 
This sustainability metric was introduced in \citet{perolat2017multi} and later used by Inequity Aversion \citep{hughes2018inequity}, Reciprocity via motivation to imitate \citep{eccles2019learning} and Gifting \citep{lupu2020gifting}.

\textbf{Equality} 
The Gini coefficient is commonly used to measure the inequality of achieved rewards within the system:

\begin{equation}
G = \frac{\sum_{i=1}^{n} \sum_{j=1}^{n} | U_i - U_j |}{2n^2 \bar{U}}.
\end{equation}

The metric $E := 1 - G$ of equality is used by the three works mentioned above alongside sustainability.

\textbf{Mechanism-based metrics}
Given a novel methodological mechanism that has been introduced, a common evaluative technique is to measure how well that mechanistic component has done its job. For example, Gifting introduces the mechanism of allowing agents to gift rewards to others and accordingly, one evaluation metric used is the gifting action frequency. Another example is Learning to Incentivise Others (LIO) \citep{yang2020learning} which introduces a similar mechanism to allow agents to provide incentives to other agents, which can also be used as metrics for evaluation.
% Here again, received incentives is a metric used for evaluation.\yali{i need to check}
Classifier Norm Model (CNM) \citep{vinitsky2023learning} used the fact that agents can sanction or "zap" other agents and measured this by monitoring the evolution of the zap likelihood. In Social Influence \citep{jaques2019social}, there are influencer agents who emit symbols before performing actions in order to influence others. The authors introduce some custom metrics to evaluate this mechanism. Speaker consistency measures how consistently an influencer emits a particular symbol before performing a particular action and instantaneous coordination measures the mutual information between a) the influencer's symbol and the influencee's next action and b) the influencer's action and the influencee's next action.

\textbf{Task-based metrics}
Many metrics measure the accomplishment of task-specific goals assumed to relate to enhancing cooperativeness within the particular task. Examples of this include the waste cleared metric used for Clean-up by Inequity Aversion \citep{hughes2018inequity} and Reciprocity \citep{eccles2019learning} via motivation to imitate \citep{eccles2019learning}, apple consumption and berry fraction evolution used respectively by Inequity Aversion and CNM \citep{vinitsky2023learning} for their fruit-gathering based tasks, and proportion of own coins collected, used by Reciprocity and LOLA \citep{foerster2018learning} for their Coin Game experiments. Social Value Orientation (SVO) \citep{mckee2020social} analyses abstention from depleting resources (a task-specific measure of sustainability) and interagent distance, measuring how well agents divide up the map between themselves. Finally, in Reputation \citep{mckee2021deep}, there are metrics measuring group contribution, territoriality and turn-taking, each assessing cooperative behaviours relevant to Clean-up.

\section{Conclusion and Future Directions  }\label{sec:conclusion}
In this paper, we have presented a comprehensive exploration of cooperation within the realm of multi-agent learning. An exhaustive review has been conducted, scrutinising contemporary advancements in cooperation across diverse scenarios — be it unified payoffs or distinct individual payoffs, and spanning both centralised and decentralised training paradigms. Furthermore, we proffer an encompassing compilation of benchmarks tailored explicitly for the evaluation of cooperative multi-agent learning endeavors. Conclusively, we delineate prospective research trajectories and underscore extant gaps in the literature that warrant future scholarly attention.
Below we suggest some areas we believe would be highly beneficial for future research based on the overview we have presented. Specifically, we highlight possible future advances in emergent cooperation, generalisation and evaluation techniques.

\paragraph{Foundation models as Agents}
Recently, large language models (LLMs) have demonstrated remarkable
potential in achieving human-level intelligence through the acquisition of vast amounts of web knowledge. This has sparked an upsurge in studies investigating LLM-based autonomous agents. 
In the context of multi-agent cooperation, we are interested in promote cooperation harnessing LLMs.
Early attempts \citep{hong2023metagpt,zhang2023proagent} show that these models often perform better when prompted with one specific objective and relevant context than with a larger range of goals and sub-goals at once.
% ooperative multi-agent structures  for how to have the large language models produce something like a larger software project. 
% This motivated by the fact that these models often perform better when prompted with one specific objective and relevant context than with a larger range of goals and sub-goals at once.
The techniques of retrieving relevant context and individual prompting have also been used to create social simulations where a larger number of players go about their daily lives in a simulated world \citep{park2023generative, vezhnevets2023generative}. The potential for studying multi-agent cooperation in such settings is still not fully understood though it likely to be substantial.

\paragraph{\text{Cooperation with Novel Agents (zero-shot social generalization)}}
Zero-shot generalization is a setting where there is a relative dearth of dedicated RL-based methods. We refer here to environments where agents must learn to cooperate with heterogeneous other agents and achieve good zero-shot cooperation performance with strangers at test time. In this case the demand is for agents and populations of agents capable of functioning both as visitors to an unfamiliar culture not seen during training, as well as in the role of the dominant ``resident'' population---joined by visitors from outside who were never encountered during training. The two cases are quite different. When an agent visits a larger unfamiliar group it often must adapt to the local conventions, which may be unfamiliar. When a population of agents are resident (they are the majority), then it is up to them to provide public goods and resolve social dilemmas, and to do so while remaining robust to distraction from new joiners. Melting Pot \citep{leibo2021scalable, agapiou2022melting} is a large suite of environments specifically for studying zero-shot generalization, broadly construed. Most environments in Melting Pot are mixed motive though Melting Pot also includes some pure common-interest environments (Overcooked) and some team-based zero-sum environments (capture the flag and king of the hill).

In the case of test-time agents being humans or based on human data, there is a scarcity in adequate evaluation benchmarks, given that it can be rather cumbersome to obtain human or human data-based heterogeneous agents. PECAN \citep{lou2023pecan} is a good example of this very approach evaluated on Overcooked, but more such environment suites and algorithms are needed. One possible idea is a tournament-style evaluation of algorithms amidst participating human agents, such as was done in \citet{axelrod80__effective_choice_in_the_prisoners_dilemma} for Iterated Prisoner's Dilemma.

% ? What distribution?

% {\rc 
% \citep{jaderberg2019human} used PBT, evolution in PBT.

% would self-play improve capacity of cooperation? 

% transitivity issues? 

% Generalise to cooperative games

% self-play drive the capacity of cooperation with different type of players. }

% \yali{try to understand what could we do with this things, and how to proper evlaute.}

\paragraph{\text{Evaluation}}
% \textcolor{red}{TODO: say something about melting pot here in relation to the massive multiplayer world ideas to develop in this paragraph.} \textcolor{red}{There are probably things to say about frontiers of cooperative MARL evaluation here too.}

Most evaluation benchmarks in cooperative MARL, especially in mixed-motive settings, tend to be toy environments, e.g. gridworld-based. Some team game environments like SMAC and Google Research Football are more complex and realistic but more work is needed to model this realism when individual incentives may not be aligned to the group incentive. Future work could focus on domains like autonomous driving and robotics, where realistic simulators are already available. In these cases, significant strategic complexity could be added through e.g. road map design, social vehicle heterogeneity or large systems of cooperating warehouse robots.
We also identified a gap in the field concerning the development of standardised evaluation metrics specifically designed to measure cooperativeness.
As discussed, sustainability and equality are two existing examples, but more are needed. This is evident in the fact that most works introduce highly specific task- and mechanism-based metrics. A new set of generally applicable metrics covering all the core concepts of the currently used specific metrics would be of great value to the community and would allow standardised benchmarking of newly introduced techniques.

{

% Large transformer models \citep{vaswani2017attention} have in recent time shown very strong results, in particular as large language models \citep{brown2020} but also for robotics \citep{brohan2023rt1} trained by reinforcement learning for wide ranges of tasks in simulated physical environment showing an ability to quickly utilise demonstrated or discovered skills using in-context learning. 

% As the large language models have been trained on very large amounts of human text, a question is what cooperative and competitive behaviours have they picked up. One early attempt \citep{akata2023playing} to evaluate this was to set them up playing repeating games like iterated prisoners dilemmas. 

% Another quite different form of multi-agent work in this sphere has been setting up cooperative multi-agent structures \citep{hong2023metagpt} for how to have the large language models produce something like a larger software project. This motivated by the fact that these models often perform better when prompted with one specific objective and relevant context than with a larger range of goals and sub-goals at once. The techniques of retrieving relevant context and individual prompting has also used for creating social simulations where a larger number of player character goes about their daily lives in a simulated world \citep{park2023generative}.
}
\bibliographystyle{unsrtnat}
\bibliography{main}  %%% Uncomment this line and comment out the ``thebibliography'' section below to use the external .bib file (using bibtex) .

% \input{appendix}
% possible venues for submission,  AI Journal, JMLR, ACM Computing survey, TPAMI
\end{document}